\documentclass[showpacs,pre,amsmath,amssymb,floats,twocolumn]{revtex4} 
\usepackage{colordvi,epsfig,bm,amsmath,amssymb}
\newcommand{\bbbN}      {{\mathbb{N}}}                 

\newcommand{\rd} {\mathrm d}
\newcommand{\re} {\mathrm e}

\newcommand{\vg} {{\bm g}}

\newcommand{\vp} {{\bm p}}

\newcommand{\nn}{\nonumber} 
\newcommand{\be}{\begin{equation}} 
\newcommand{\ee}{\end{equation}}  
\newcommand{\bea}{\begin{eqnarray}}
\newcommand{\eea}{\end{eqnarray}}
\newcommand{\barr}{\begin{array}}
\newcommand{\earr}{\end{array}}
\newcommand{\bcent}{\begin{center}} 
\newcommand{\ecent}{\end{center}}  

\begin{document}
\title{Heterogeneity in Outcomes of Repeated Instances of\\ Percolation Experiments }
\author{Reimer K\"uhn$^1$ and Jort van Mourik$^2$}
\affiliation{$^1$Mathematics Department, King's College London, Strand, London WC2R 2LS,UK\\
$^2$NCRG, Aston University, Aston Triangle,  Birmingham B4 7ET, UK}
\date{August 18,2020}
\begin{abstract}
\noindent
We investigate the heterogeneity of outcomes of repeated instances of percolation experiments in complex networks using a message passing approach to evaluate heterogeneous, node dependent probabilities of belonging to the giant or percolating cluster, i.e.\, the set of mutually connected nodes whose size scales linearly with the size of the system. We evaluate these both for large finite single instances, and for synthetic networks in the configuration model class in the thermodynamic limit. For the latter, we consider both Erd\H{o}s-R\'enyi and scale free networks as examples of networks with narrow and broad degree distributions respectively. For real-world networks we use an undirected version of a Gnutella peer-to-peer file-sharing network with $N=62,568$ nodes as an example. We derive the theory for multiple instances of both uncorrelated and correlated percolation processes. For the uncorrelated case, we also obtain a closed form approximation for the large mean degree limit of Erd\H{o}s-R\'enyi networks.
\end{abstract}


\maketitle

\section{Introduction}\label{sec:Intro}

Ever since the start of research into random graphs and complex networks, the problem of percolation has taken center stage, starting with determining the conditions under which random graph ensembles do exhibit a so-called giant or percolating cluster that occupies a finite fraction of the system in the large system limit \cite{ErdRen59, ErdRen60, MollReed95, MollReed98}. With the growing importance of networks and network based technologies in real life, percolation as a process on existing networks, where edges (or nodes) are kept with some probability $p$ and are deleted with probability $1-p$, has been much studied. The survival (and size) of a giant cluster is taken as a measure of the resilience of a network against random failure of nodes or links  \cite{Call+00,Cohen+00}. In these papers, generating function methods were used to evaluate the average fraction of nodes in the giant component as well as average sizes of finite connected clusters that are not part of the giant component. Such methods have also been used to analyse the sizes of avalanches of cascading failures in interacting systems \cite{Watts02}. Studies of percolation in complex networks, both with and without additional structure have been linked to the issue of network resilience against random failure or intentional disruption of components  ever since; for some more recent results see e.g. \cite{Braunstein+PNAS16, Bianconi18, Wandelt+SciRep18, Dong+PNAS18,Tishby+18b, AllardDufr19, DufrAllard19,Baxter+20} and references therein. However, it is also worth highlighting another important aspect of network resilience that goes {\em beyond\/} connectivity properties as captured by percolation, namely the issue of the integrity of non-trivial collective states in networked systems with interacting degrees of freedom; see e.g. \cite{AnandKuehn07, Nasiev+07, Hase08, Annibale+10}. Indeed, while stochastic dynamical systems defined on complex networks could not support non-trivial collective states without a giant connected component, such states may become unstable as a result of random (or targeted) removal of links or nodes well before the giant component disappears.

There is an interesting link between the long term behaviour of SIR (susceptible-infected-recovered) models of infection dynamics and bond percolation on complex networks, which appears to have been made as early as 1983 \cite{Grassb83}. It was generalized to cover heterogeneous  transmission processes \cite{Sand+02}, and investigated using generating function methods in \cite{New02}, concentrating on instabilities against outbreaks and on the average size of the epidemic. Studies of the {\em dynamics\/} of epidemics in complex networks on the other hand, rather than concentrating on overall average probabilities of infection or recovery, have resorted to a heterogeneous dynamic mean-field theory which allows one to take (some of) the heterogeneity of network structures into account \cite{PastorVesp01, More+02}. In these studies a so-called degree based approximation is adopted which assumes that the fate of a node in a complex network during an epidemic depends {\em solely\/} on its degree. While this simplifies matters sufficiently for equations of motion to be analytically tractable, it misses several key aspects of the full heterogeneity in the problem. The design of optimal immunization strategies using degree based heterogeneous mean-field theories \cite{ Cohen+03, Madar+04} may therefore well miss opportunities as further aspects of heterogeneity could be exploited. 
For a recent overview we refer to \cite{Pastor+15}. 

The formulation of a cavity or message passing approach to single network instances of percolation probabilities \cite{ShirKaba10, Karrer+14} paved the way to assess the fate of individual nodes under random bond (or site) removal. This was initially investigated to some extent in the context of infection dynamics in \cite{Rog15} with the full heterogeneity in the problem first exposed only in \cite{KuRog17}. It is worth noting that degree based information {\em can\/} also be recovered from the generating function approach, which has traditionally been used to analyse mainly average behaviour. Indeed, in \cite{Tishby+18} it was demonstrated that degree based information about percolation probabilities can be obtained using expressions for the average percolation probabilities by `unfolding' them according to degree. In that paper it was also shown that iterated versions of the self-consistency equations from which percolation probabilities are normally obtained, can be used to go beyond degree based approximations and recover the full distributions of percolation probabilities first obtained in \cite{KuRog17}.

The distribution of percolation probabilities of individual nodes in complex networks is one aspect of the variability in the percolation problem one may want to characterize. Another aspect that was recently addressed is the question of {\em  fluctuations\/} of percolation probabilities  of network nodes across two separate random (not necessarily independent) realizations of percolation experiments \cite{Bianconi17}. A closely related problem concerns the evaluation of {\em joint\/} percolation probabilities, more specifically the question of  determining the fraction of nodes that would be part of the giant cluster in all of a set of $\tau$ (once more not necessarily independent) percolation experiments, which was recently solved in \cite{Kitsak+18}, and has been discussed as a measure of the stability of the giant cluster in the percolation problem.  In \cite{Bianconi17, Kitsak+18} these questions were mostly addressed at a global level, though \cite{Kitsak+18} have looked at local signatures, such as the influence of node degrees (as predicted within a generating function approach), and used simulations to assess heterogeneity of degree dependent outcomes. 

In the present paper, we take into account the full heterogeneity of the problems studied in \cite{Bianconi17, Kitsak+18}. For the sake of definiteness we consider bond percolation. However, the methods can easily be adapted to cover node percolation as well. Our paper is organized as follows. In order to keep the paper self-contained, we briefly review in Sect.~\ref{sec:Perc} the well-known message-passing approach to bond percolation, starting with the formulation for large single instances, and then formulating equations describing the limit of infinite system size for networks in the configuration model class. In Sect.~\ref{sec:FluctCorr} we then formulate the theory for multiple instances of the percolation processes, starting with independent instances, which will allow us to uncover the full heterogeneity in the problem of the stability of the giant cluster studied in \cite{Kitsak+18}. In Sect.~\ref{sec:FluctCorr}.B we derive closed form expressions for the distribution of percolation probabilities for multiple uncorrelated instances in the large mean degree limit for Erd\H{o}s-R\'enyi (ER) networks. In Sect.~\ref{sec:FluctCorr}.C we present the modifications required to cover the effect of {\em correlation\/} between multiple instances of the percolation processes. This allows us to analyze the full heterogeneity of the fluctuation problem studied in \cite{Bianconi17}, and to generalize it beyond the two-instances case. Sect.~\ref{sec:FluctCorr}.D analyses more general correlations. Our main results are presented and discussed in Sect.~\ref{sec:Res}, and we summarize and discuss our findings in Sect.~\ref{sec:SumDisc}.

At this point it is useful to compare and contrast the nature of the heterogeneity treated in the present paper with other forms of variability that have been discussed in the context of percolation. The size of the giant connected component  (measured as a fraction of system size) has been proven to be self-averaging in the thermodynamic limit in networks of the configuration model class \cite{BallNeal17}. Although a formal proof is still missing, the same is expected to hold for the distribution of node-dependent percolation probabilities studied in \cite{KuRog17}. The origin of the heterogeneity in that problem is indeed very simple and related to the fact that upon random node or link removal, nodes with high connectivity to the densest regions of a network will continue to have a high probability to remain part of any giant connected component, while nodes whose connection to dense regions of a network is tenuous will have low probability to do so. 
In essence, the variability of local environments creates the heterogeneity of percolation probabilities for any typical realization of a percolation process. This type of heterogeneity may make it difficult to properly identify the emergence of a giant percolating component in some real world networks of (fixed) finite size \cite{DufrAllard19}. Indeed, in networks of finite size, {\em every\/} connected component occupies a finite fraction of the system. Therefore, traditional methods for finding the  location of the incipient percolation transition may give conflicting results \cite{DufrAllard19}, and cannot be resolved by a finite-size scaling analysis, as the size of any given real world network cannot be varied.

The type of heterogeneity just described is radically different from the (dynamical) heterogeneity that gives rise to non self-averaging time-dependent overall percolation probabilities in so-called explosive percolation \cite{Achliop+Sci09}. The phenomenology is observed in network growth processes with ``choice". Rather than randomly linking up pairs of nodes as in the ER model \cite{ErdRen59,ErdRen60}, more than one random edge is proposed to be connected to the growing network, but only one of them is actually selected in a manner that is designed to delay the emergence of a giant connected component. This can result in a (delayed) explosive percolation phenomenon, for which the time dependent fraction of nodes in the giant component can in some cases be shown to be non self-averaging, even in the thermodynamic limit \cite{RiordanWarnke12}; for a recent review we refer to \cite{ExplPerc19}. 

Another form of heterogeneity is related to {\em rare\/} realizations of the configurations of links or nodes that are actually removed from the system in a percolation process. For ER networks, they were first studied using Large Deviations theory in \cite{Engel+04}. In a subsequent numerical study \cite{Hartmann11} of percolation on ER networks and 2d lattices it was shown that such rare events can give rise to bi-stability or coexistence of non-percolating and percolating configurations in large finite systems. This phenomenon was recently confirmed in \cite{Bianconi18, Coghi+18}, using a combination of a message passing approach and Large Deviations techniques. Rare configurations of removed nodes where shown to be able to suppress the giant component for values of the node removal probability where it would exist if configurations of node removals were {\em typical\/}.

\section{Bond Percolation}\label{sec:Perc}
We consider a percolation process on graphs in the configuration model class. To investigate the heterogeneity of outcomes of repeated instances of percolation experiments on the same graph, we use a message passing approach \cite{ShirKaba10,Karrer+14}. We follow the methods outlined in \cite{KuRog17} to expose the heterogeneity in the results. The approach taken in \cite{Karrer+14} investigates percolation by focusing on distributions of sizes of finite clusters. However, as we concentrate solely on percolation probabilities here, we adopt a slightly different approach from that in \cite{Karrer+14, KuRog17}, using a message passing formulation closer to that of \cite{ShirKaba10} to directly determine the probability for a given node to belong to the giant or percolating component (abbreviated by GC from now on), if it exists.

Networks in the configuration model class are maximally random, subject to a prescribed degree distribution. Thus, denoting by $k_i$ the degree of node $i$, one has that $p_k = {\rm Prob}(k_i = k)$ for some degree distribution $\vp = (p_k)_{k\in\bbbN}$, and there are no degree-degree correlations.

\subsection{Message passing for Large Single Network Instances}
To formulate the message passing approach, we introduce indicator variables $n_i$ denoting whether node $i$ is in the GC ($n_i=1$) or not ($n_i=0$), and link variables $x_{ij}$ that denote whether the edge $(ij)$ is kept in a single realization of the percolation process ($x_{ij}=1$) or not ($x_{ij}=0$). Clearly, for a node $i$ to be in the GC it must be connected to it through at least one of its neighbours. This requires, for at least one of the edges $(ij)$ connecting to node $i$, that both the edge is kept in the percolation process ($x_{ij}=1$), {\em and\/} that the node $j$ is itself in the GC, even on a graph from which node $i$ and all edges emanating from it are removed. Such a graph is usually referred to as a cavity graph. Introducing an indicator variable $n_j^{(i)}$, taking the value 1, if node $j$ neighbouring on $i$ is indeed in the GC on the cavity graph, and 0 if it is not, the condition above is expressed as
\be
n_i = 1 - \prod_{j\in\partial i}
\Big(1 -x_{ij} n_j^{(i)}\Big)\ ,
\label{ni-eq}
\ee
where $\partial i$ denotes the set of nodes connected to $i$ on the original graph.\\
For the (cavity) indicator variables  $n_j^{(i)}$ we have  by analogous reasoning
\be
n_j^{(i)} = 1 - \prod_{\ell\in\partial j\setminus i}\Big(1 -x_{j\ell} n_\ell^{(j)}\Big)\ ,
\label{nji-eq}
\ee
where $\partial j\setminus i$ denotes the set of nodes connected to $j$ on the cavity graph with $i$ removed.

Averaging Eqs.\,(\ref{ni-eq}) over all realizations of the percolation process gives
\be
g_i = 1 - \prod_{j\in\partial i}\Big(1 - p g_j^{(i)}\Big)
\label{avni-eq}
\ee
for the probability $g_i$ for node $i$ to belong to the GC under the percolation process. This result, assumes independence of the random variables associated with different edges emanating from a node, and is only exact on trees for which averages over different branches factor. It is generally assumed (and confirmed by experiments) that this is an excellent approximation for large finitely connected systems, which are locally tree-like, and that it becomes asymptotically exact in the limit of infinite system size $N\to \infty$. Following the same logic, averaging \eqref{nji-eq} gives
\be
g_j^{(i)} = 1 - \prod_{\ell\in\partial j\setminus i}\Big(1 -p g_\ell^{(j)}\Big)
\label{avnji-eq}
\ee
for the probability of node $j$ adjacent to $i$ to be part of the giant cluster on the cavity graph with $i$ removed.

Equations (\ref{avnji-eq}) can be solved iteratively on any large single instance of a graph, and the site dependent percolation probabilities $g_i$ can then be computed using Eq.\,\eqref{avni-eq}. Note that as the solutions to Eqs.\,(\ref{avnji-eq}) will be heterogeneous due to their local environments, so will the $g_i$ even across nodes with the same degree.

\subsection{Thermodynamic Limit}
In the thermodynamic limit Eqs.\,(\ref{avnji-eq}) constitute an infinite recursion. Assuming that a probability law exists for the $g_j^{(i)}$, the probability density $\tilde \pi(\tilde g)$ can be obtained by demanding probabilistic consistency of Eqs.\,(\ref{avnji-eq}). Following by now standard reasoning, $\tilde \pi(\tilde g)$ is obtained by summing probabilities of all realizations of the r.h.s. of Eqs.\,(\ref{avnji-eq}) for which $g_j^{(i)}\in (\tilde g,\tilde g+ \rd\tilde g]$, assuming that the $g_\ell^{(j)}$ on the r.h.s. in Eqs.\,(\ref{avnji-eq}) are drawn independently from  $\tilde \pi$. Hence, Eqs.\, (\ref{avnji-eq}) translate into 
\be
\tilde \pi(\tilde g) = \sum_k \frac{k}{c}\,p_k \, \tilde\pi(\tilde g|k)
\label{pit-gt}
\ee
with
\be
\tilde \pi(\tilde g|k) = \int \Big[\prod_{\nu=1}^{k-1} \rd \tilde 
\pi(\tilde g_\nu)\Big]\, \delta\Big(\tilde g - \Big(1-\prod_{\nu=1}^{k-1} (1-p \tilde g_\nu)\Big)\Big)\ .
\label{pit-gtk}
\ee
Here $\frac{k}{c}\,p_k$ is the probability that a randomly chosen edge links to a node of degree $k$, and $\delta(\cdot)$ is the Dirac $\delta$-distribution. Furthermore, we have adopted the shorthand $ \rd \tilde \pi(\tilde g_\nu) =  \rd \tilde  g_\nu\, \tilde \pi(\tilde g_\nu)$.

Equation (\ref{pit-gt}) is efficiently solved using a population dynamics algorithm. The distribution $\pi(g)$ of node dependent percolation probabilities $g_i$ is then similarly obtained  from Eq.\,\eqref{avni-eq}, 
to give
\be
\pi(g) = \sum_k p_k \pi(g|k)\ ,
\label{pi-g}
\ee
in which the
\be
\pi(g|k) = \int \Big[\prod_{\nu=1}^{k} \rd \tilde \pi(\tilde g_\nu)\Big] \,
\delta\Big(g - \Big(1-\prod_{\nu=1}^{k} (1-p \tilde g_\nu)\Big)\Big)
\label{pi-gk}
\ee
are the distributions percolation probabilities {\em conditioned\/} on nodes having degree $k$. It is straightforward to show that these equations are equivalent to the marginal densities describing the percolation-probability sector in \cite{KuRog17}, which were obtained following a different route based on cluster-size distributions.

\section{Fluctuations, Correlations, and Stability of the Giant Cluster}\label{sec:FluctCorr}

We now turn to the fluctuations and correlations of percolation probabilities, and the stability of the GC under repeated instances of a bond percolation process. The fluctuations and correlations of percolation probabilities were recently investigated at the {\em global} level in \cite{Bianconi17}, whereas the stability of the GC was investigated in \cite{Kitsak+18}, again mostly at the {\em global} level.

We start our study of the {\em local\/} statistics of these quantities with the case of independent instances of the percolation process, thereafter generalizing to correlated instances.

\subsection{Independent Instances}
Let $n_i(\tau)$ denote the indicator variable that designates whether node $i$ is part of the GC for {\em all} instances in a set $\tau$ of percolation experiments ($n_i(\tau) = 1$), or not  ($n_i(\tau) = 0$). Labeling individual instances by $t\in\tau$, from (\ref{ni-eq}) we obtain $n_i(\tau) = \prod_{t\in \tau} n_i(t) $, so
\be
n_i(\tau)  = \prod_{t\in \tau} \Big( 1 - \prod_{j\in\partial i}\Big(1 -x_{ij}(t) n_j^{(i)}(t)\Big)\Big)
\label{ni-tau}
\ee
where $x_{ij}(t)$ indicates whether the edge $(ij)$ is present in experiment $t$, $(x_{ij}(t)=1)$, or not $(x_{ij}(t)=0)$. Similarly $n_j^{(i)}(t)\in\{1,0\}$ denotes whether or not node $j$ is in the GC of the cavity graph (with node $i$ removed) in instance $t$ . Clearly,
\be
n_j^{(i)}(t) =  1 - \prod_{\ell\in\partial j\setminus i}\Big(1 -x_{j\ell}(t) n_\ell^{(j)}(t)\Big)
\label{nji-t}
\ee
 as in Eq. (\ref{nji-eq}).

When the percolation experiments are independent, Eqs. \eqref{ni-tau} and \eqref{nji-t} can be straightforwardly averaged giving
\be
g_i(\tau) =\langle n_i(\tau) \rangle= \prod_{t\in \tau} \Big( 1 - \prod_{j\in\partial i}\Big(1 -p_t g_j^{(i)}(t)\Big)\Big)
\label{gi-tau}
\ee
for the probability of node $i$ to belong to the GC for all independent instances in the set $\tau$. Here $p_t =\langle x_{ij}(t)\rangle$ is the overall probability to retain bonds in percolation experiment $t$. The cavity probabilities $g_j^{(i)}(t)$ must satisfy
\be
g_j^{(i)}(t) = 1 - \prod_{\ell\in\partial j\setminus i}\Big(1 - p_t g_\ell^{(j)}(t)\Big)\ ,
\label{gji-t}
\ee
and are independent of $t$ if the bond retention probabilities $p_t$ are. Henceforth, we assume that $p_t\equiv p,\quad\forall t\in\tau$, though it is clear that generalizing to experiment-dependent edge retention probabilities is straightforward.

The pdf $\pi_\tau$ of the joint percolation probabilities $g_i(\tau)$ is then obtained exploiting the independence of the $g_j^{(i)}(t)$. Thus, by the same line of reasoning that led to Eqs.\,\eqref{pi-g},\eqref{pi-gk}, we obtain
\be
\pi_\tau(g) = \sum_k p_k \pi_\tau(g|k)
\label{pitau-g}
\ee
in which the
\be
\pi_\tau(g|k) = \int \Big[\prod_{\nu=1}^{k} \rd \tilde \pi(\tilde g_{\nu})\Big] \,
\delta\Big(g - \Big(1-\prod_{\nu=1}^{k} (1-p \tilde g_{\nu})\Big)^{|\tau|}\Big)\ \ ,
\label{pitau-gk}
\ee
are the distributions of joint percolation probabilities {\em conditioned\/} on nodes having degree $k$, and where $|\tau|$ denotes the size of the set $\tau$ and $\tilde \pi$ is a solution of \eqref{pit-gt}. For the {\em average\/} probability
\be
\langle g(\tau) \rangle = \int \rd \pi_\tau(g)\,  g
\ee
for a node to belong to the giant component in a set $\tau$ of independent percolation experiments, we then obtain
\be
\langle g(\tau)\rangle  =\sum_k p_k \Big(1 - (1-p \langle \tilde g\rangle)^k\Big)^{|\tau|}\ ,
\ee
where $\langle\tilde g\rangle =\int\rd\tilde g \tilde\pi(\tilde g)\tilde g$. Note that this result was obtained directly by considering average behaviour in \cite{Kitsak+18}.

\subsection{The Large Mean Degree Limit}
For `narrow' degree distributions where the standard deviation of the degrees is much smaller than the mean degree, it is relatively straightforward \cite{KuRog17} to obtain closed form approximations of the results above. Here we consider the Poisson degree distribution of ER graphs with large mean degree $\langle k\rangle = c$, for which the standard deviation $\sigma_k=\sqrt{c}$ is small compared to the mean for $c\gg 1$.

In the large mean degree limit the solution of \eqref{pit-gt} is well approximated by the $\delta$-distribution $\tilde\pi(\tilde g) = \delta(\tilde g - \tilde g_*)$. The value of $g_*$ is obtained by inserting this ansatz into \eqref{pit-gt}, and deriving a self-consistency equation for $g_*$. Assuming a Poisson distribution for the degrees, we get
\be
 g_* = 1 - \re^{-p c g_*}\ .
\label{gstar}
\ee
In order to obtain a non-trivial solution in the large $c$ limit, one has to adopt the scaling $p = \rho/c$ at fixed $\rho$, so that \eqref{gstar} becomes
\be
 g_* = 1 - \re^{-\rho g_*}\ ,
\label{gstarn}
\ee
which can be solved in closed form, giving
\be
g_* =  1 + \frac{W(-\rho \re^{-\rho})}{\rho}\ ,
\ee
where $W$ is the Lambert $W$-function. For real valued arguments $x \ge -\re^{-1}$, its value $W(x)$ is defined as (the principal branch of) the solution of the transcendental equation $W \re^W = x$; see Sect.\,4.13 in \cite{NIST:DLMF}.

In order to obtain the large mean degree limit of the distribution $\pi_\tau$ of joint percolation probabilities for $|\tau|$ independent percolation experiments, we insert these results into Eq. \eqref{pitau-g} that implies that the conditional probability for a node of degree $k$ to belong to the giant cluster is
\be
g = g_\tau(k) = \big[1 - (1 - p g_*)^k\big]^{|\tau|} 
\label{gtauofk}
\ee
For a Poisson distribution of  large mean degree, $c \gg 1$, the distribution of scaled degrees $x = k/c$ is well approximated by a normal distribution of mean 1 and variance $1/c$. Hence from \eqref{gtauofk}, we derive a closed form expression for the pdf $\pi_{\tau}(g)$ as follows. From \eqref{gtauofk}, we have
\be
x =x(g) = \frac{\log(1 - g^{1/|\tau|})}{c \log(1 - p g_*)} \ ,
\ee
such that a normal distribution $\pi(x) = \sqrt{\frac{c}{2\pi}}\,\exp\big[- \frac{c}{2} (x-1)^2\big]$ transforms into
\bea
~\hspace{-6mm}\pi_\tau(g) &=&\pi(x)\Big|\frac{\rd x}{\rd g}\Big| \nn\\
&=& - \frac{1}{\sqrt{2\pi c}}\,
\frac{g^{-(1-1/|\tau|)}}{|\tau|\, (1- g^{1/|\tau|})\,\log(1 - p g_*)}\nn\\
& &  \times \exp\Bigg[-\frac{c}{2}
\left(\frac{\log(1 - g^{1/|\tau|})} {c \log(1 - p g_*)} -1\right)^2\Bigg]\ .
\label{piofg-largec}
\eea

In Sec.\ref{sec:Res} we show that even for moderate values of the mean degree $c$ this already provides a decent approximation for the distribution of percolation probabilities across several independent percolation experiments.

\subsection{Correlated Instances}

When the instances of the percolation process are {\em not independent\/}, the analysis becomes more involved. To average $n_i(\tau)$ in Eq.\,\eqref{nji-t} over the joint distribution of the instances for all $t\in \tau$, we first expand the products appearing in \eqref{nji-t}:
\bea
n_i(\tau) &=& \sum_{\sigma\subseteq \tau} (-)^{|\sigma|} \prod_{j\in\partial i}\Bigg[ \sum_{\sigma'\subseteq \sigma}  (-)^{|\sigma'|}\nn\\
& &\hspace{12mm}\times \prod_{t\in\sigma'} \Big(x_{ij}(t) n_j^{(i)}(t)\Big)\Bigg]\ .
\label{ni-tau-exp}
\eea
The averages on the r.h.s. do factor w.r.t. $j$ due to the assumed locally tree-like nature of the systems we consider. However, averages over the $t$-products do {\em not\/} factor w.r.t. $t$,  although averages of the form $\big\langle \prod_t \big(x_{ij}(t) n_j^{(i)}(t)\big) \big\rangle$ decouple in the $x_{ij}$- and $n_j^{(i)}$-sectors:
\be
\Big\langle \prod_{t\in \sigma} \Big( x_{ij}(t) n_j^{(i)}(t)\Big)\Big \rangle = \Big\langle \prod_{t\in \sigma} x_{ij}(t)\Big  \rangle
\Big\langle \prod_{t\in \sigma} n_j^{(i)}(t)\Big \rangle \ .
\ee
Assuming the statistics of the $x_{ij}(t)$ to be uniform and independent across edges $(ij)$, and introducing
\be
p(\sigma) = \Big\langle \prod_{t\in \sigma} x_{ij}(t)\Big  \rangle
\label{psig}
\ee
and
\be
g_j^{(i)}(\sigma)=\big \langle n_j^{(i)}(\sigma) \big\rangle = \Big\langle \prod_{t\in \sigma} n_j^{(i)}(t)\Big \rangle \ ,
\label{gtsig}
\ee
we obtain $g_i(\tau) = \langle n_i(\tau)\rangle$, giving
\be
g_i(\tau)  =  \sum_{\sigma\subseteq \tau} (-)^{|\sigma|} \prod_{j\in\partial i}
\Bigg[ \sum_{\sigma'\subseteq \sigma}  (-)^{|\sigma'|} p(\sigma') g_j^{(i)}(\sigma')\Bigg]\ ,
\ee
with the convention $p(\emptyset)=g_i(\emptyset)=g_j^{(i)}(\emptyset) =1$.

An entirely analogous line of reasoning, for any $\rho \subseteq \tau$ yields a set of self-consistency equations for the cavity expectations
\be
g_j^{(i)} (\rho) =  \sum_{\sigma\subseteq \rho} (-)^{|\sigma|} \prod_{\ell\in\partial j\setminus i}
\Bigg[ \sum_{\sigma'\subseteq \sigma}  (-)^{|\sigma'|} p(\sigma') g_\ell^{(j)}(\sigma')\Bigg]\  .
\label{gji-rho}
\ee

The equations \eqref{gji-rho} for $\rho\subseteq \tau$ define a hierarchy of $2^{|\tau|}-1$ $\rho$-point functions for each of the edges $(ij)$ of the graph which are parametrized by the set $\{p(\sigma); \sigma\subseteq \tau\}$ of edge-occupancy expectations that represent the dynamical model underlying correlations of edge-occupancy.  Note that the $g_j^{(i)} (\rho)$ depend on all  $g_j^{(i)}(\sigma)$ with $\sigma\subseteq \rho$. These can in principle be solved for any given large single instance of a graph, building the hierarchy starting from the one-point functions required in the standard percolation problem, using these to solve for all two-point functions, using one-point and two-point functions to solve for all three-point functions, and so on.

Alternatively, one can formulate a self-consistency equation for the pdf $\tilde\pi$ of the $\vg_j^{(i)} =\big (g_j^{(i)} (\rho)\big)_{\rho \subseteq \tau}$. With reference to \eqref{gji-rho}, we obtain
\be
\tilde\pi(\tilde \vg) =  \sum_k \frac{k}{c}\,p_k \tilde\pi(\tilde \vg|k)\ ,
\label{tpivgt}
\ee
with
\bea
\tilde\pi(\tilde \vg|k) &=&  \int \Big[\prod_{\nu=1}^{k-1}\,
\rd \tilde\pi(\tilde \vg_\nu)\Big] \prod_{\rho\subseteq\tau}\,
\delta\Big(\tilde g_\rho -  \sum_{\sigma\subseteq \rho} (-)^{|\sigma|} \nn\\
& &\hspace{4mm}\times \prod_{\nu=1}^{k-1}
\Big[\sum_{\sigma'\subseteq \sigma}  (-)^{|\sigma'|} p(\sigma') \tilde g_\nu(\sigma')\Big] \Big)\ ,
\label{tpivgtk}
\eea
in complete analogy to constructions used earlier. Once the solution of these self-consistency equations for the $\tilde\pi$ is found, the pdf $\pi(\vg)$ of the $\vg_i=\big(g_i(\rho)\big)_{\rho\subseteq \tau}$ is then given by
\be
\pi(\vg) =  \sum_k \,p_k \pi(\vg|k)\ ,
\label{pivg}
\ee
in which the
\bea
\pi(\vg|k) &=& \int \Big[\prod_{\nu=1}^{k} \,\rd \tilde\pi (\tilde \vg_\nu)\Big]\,
 \prod_{\rho\subseteq\tau} \delta\Big(g_\rho -  \sum_{\sigma\subseteq \rho} (-)^{|\sigma|}\nn\\
& &\hspace{5mm}\times \prod_{\nu=1}^{k}
\Big[\sum_{\sigma'\subseteq \sigma}  (-)^{|\sigma'|} p(\sigma') \tilde g_\nu(\sigma')\Big] \Big)
\label{pivgk}
\eea
are now a joint distributions for a set of $2^{l\tau|-1}$ $\ell$-point functions with $1\le \ell\le|\tau|$, defined in terms of all non-empty subsets of $\tau$, {\em conditioned\/} on nodes having degree $k$.

Global averages are obtained by evaluating first moments of $\pi$ and the  $\tilde\pi$, for $\rho\subseteq \tau$.  Defining
\be
\langle g(\rho)\rangle = \int \rd \pi(\vg)\, g(\rho) \quad \mbox{and}\quad  \langle{\tilde g}(\rho)\rangle 
= \int \rd\tilde\pi(\tilde \vg)\,  \tilde g(\rho)\ ,
\ee
we obtain
\be
\langle g(\rho)\rangle = \sum_{\sigma\subseteq \rho} (-)^{|\sigma|} G_0\Big(\sum_{\sigma'\subseteq \sigma} 
 (-)^{|\sigma'|} p(\sigma') \langle{\tilde g} (\sigma') \rangle\Big)
\ee
fom Eqs.\,\eqref{pivg}, \eqref{pivgk}, with the  $ \langle{\tilde g} (\rho) \rangle$ satisfying the hierarchical set of self-consistency equations
\be
\langle \tilde g (\rho)\rangle = \sum_{\sigma\subseteq \rho} (-)^{|\sigma|} G_1\Big(\sum_{\sigma'\subseteq \sigma} 
 (-)^{|\sigma'|} p(\sigma') \langle\tilde g (\sigma')\rangle\Big)
\ee
derived from Eqs.\,\eqref{tpivgt}, \eqref{tpivgtk}. Here $G_0$ and $G_1$ are the generating functions of the degree distribution and the distribution of degrees of nodes reached by following a random link, respectively
\be
G_0(x)=\sum_k p_k\, x^k\quad \mbox{and}\quad 
G_1(x)=\sum_k \frac{k}{c}\,p_k\, x^{k-1}\ .
\ee

Note that the complexity of the analysis for correlated instances of percolation increases exponentially with the number of percolation experiments considered, and  so will quickly become prohibitively involved except for a relatively small number of instances.

\subsection{More general correlations}
If one were interested in more general correlations, such as probabilities that a node is part of the GC in all percolation instances in the set $\tau$ while {\em not\/} in any instance in a complementary set $\bar\tau$, one would have to consider
\be
n_i(\tau,\bar\tau) = \prod_{t\in\tau} n_i(t) \, \prod_{\bar t\in\bar\tau} (1-n_i(\bar t))
\ee
To evaluate averages over all realizations of these instances of the percolation process, one has to expand the expression of these products in terms of cavity indicator variables, giving
\bea
n_i(\tau,\bar\tau) &=& \sum_{\sigma\subseteq \tau} (-)^{|\sigma|} \prod_{j\in\partial i}
\Big[\sum_{\sigma'\subseteq \sigma} \sum_{\tau'\subseteq \bar\tau} (-)^{|\sigma'|+|\tau'|}\nn\\
& & \hspace{12mm} \times\prod_{t\in \sigma'\cup\tau'}
x_{ij}(t) n_j^{(i)}(t) \Big]\ .
\eea

Following the reasoning outlined above, the distributions of $g_i(\tau,\bar\tau) =\langle n_i(\tau,\bar\tau) \rangle$ are evaluated in terms of solutions of Eqs. \eqref{tpivgt}, \eqref{tpivgtk} as
\be
\pi_{\tau,\bar\tau}(g)   =  \sum_k \,p_k \pi_{\tau,\bar\tau}(g|k)
\ee
in which the
\bea
\pi_{\tau,\bar\tau}(g|k)\!\!\!  &=&\!\!\! \int \Big[\prod_{\nu=1}^{k} \,\rd \tilde\pi(\tilde\vg_\nu)\Big] \,
\delta\Big( g -  \sum_{\sigma\subseteq \tau} (-)^{|\sigma|}\nn\\
& &\hspace{-18mm}\times\prod_{\nu=1}^{k}\!\! \Big[\!\!\sum_{\sigma'\subseteq \sigma}  
\!\!\sum_{\tau'\subseteq \bar\tau}\!\! (-)^{|\sigma'|+|\tau'|}
 p(\sigma'\!\cup\!\tau') \tilde g_\nu(\sigma'\!\cup\!\tau')\Big] \Big) ,
\eea
are the relevant joint distributions of a hierarchy of $\ell$-point functions conditioned on node degree $k$, with $1\le \ell \le |\tau|+|\bar\tau|$; the $\tilde\pi$ are now defined over an enlarged space with $\tilde\vg = \big(\tilde g(\rho)\big)_{\rho\subseteq \tau\cup\bar\tau}$, and $\pi_{\tau,\bar\tau}(g)$ is the distribution of the probability for a node to belong to the GC in every instance of the set $\tau$ and in none of the set $\bar\tau$.

Global averages are obtained as first moments, giving
\bea
\big\langle g(\tau,\bar\tau\big)\rangle &=&  \sum_{\sigma\subseteq \tau} (-)^{|\sigma|} G_0\Big(\sum_{\sigma'\subseteq \sigma}  \sum_{\tau'\subseteq \bar\tau} (-)^{|\sigma'|+|\tau'|} \nn\\
& &\hspace{12mm} \times p(\sigma'\cup\tau') \big\langle\tilde g (\sigma'\cup\tau')\big\rangle \Big) ,
\label{avtautaubar}
\eea
which generalizes the global results of \cite{Bianconi17} to more than two possibly correlated instances of a percolation experiment.

\section{Results} \label{sec:Res}
In what follows, we first present results for the stability of the GC measured in terms of the probability that nodes of the network would be part of it in several instances of the percolation process as studied at the global level in \cite{Kitsak+18}. The second set of results are co-variances of nodes to be on the GC in two correlated instances of a percolation process. This problem was studied, once more on a global level, in  \cite{Bianconi17}.

\begin{figure*}[t!]
\includegraphics[width=0.475\textwidth]{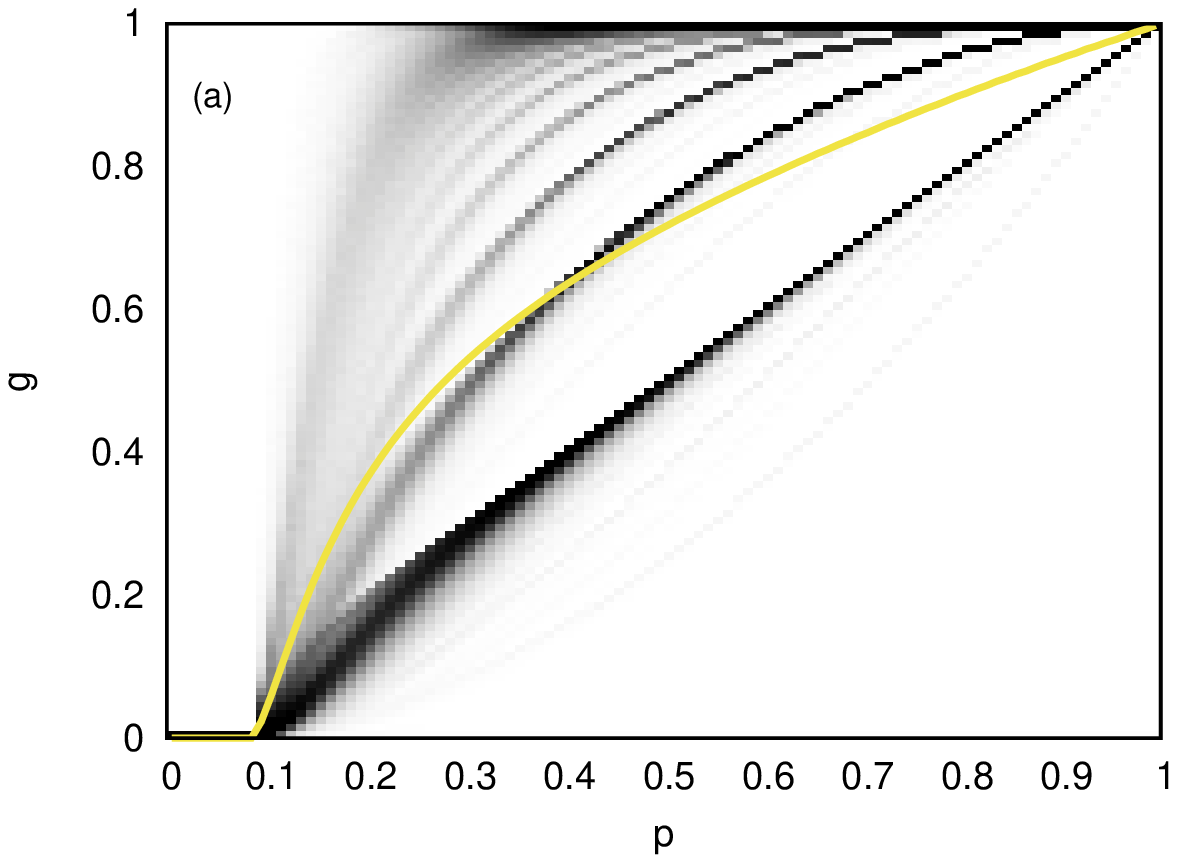}\hfil 
\includegraphics[width=0.475\textwidth]{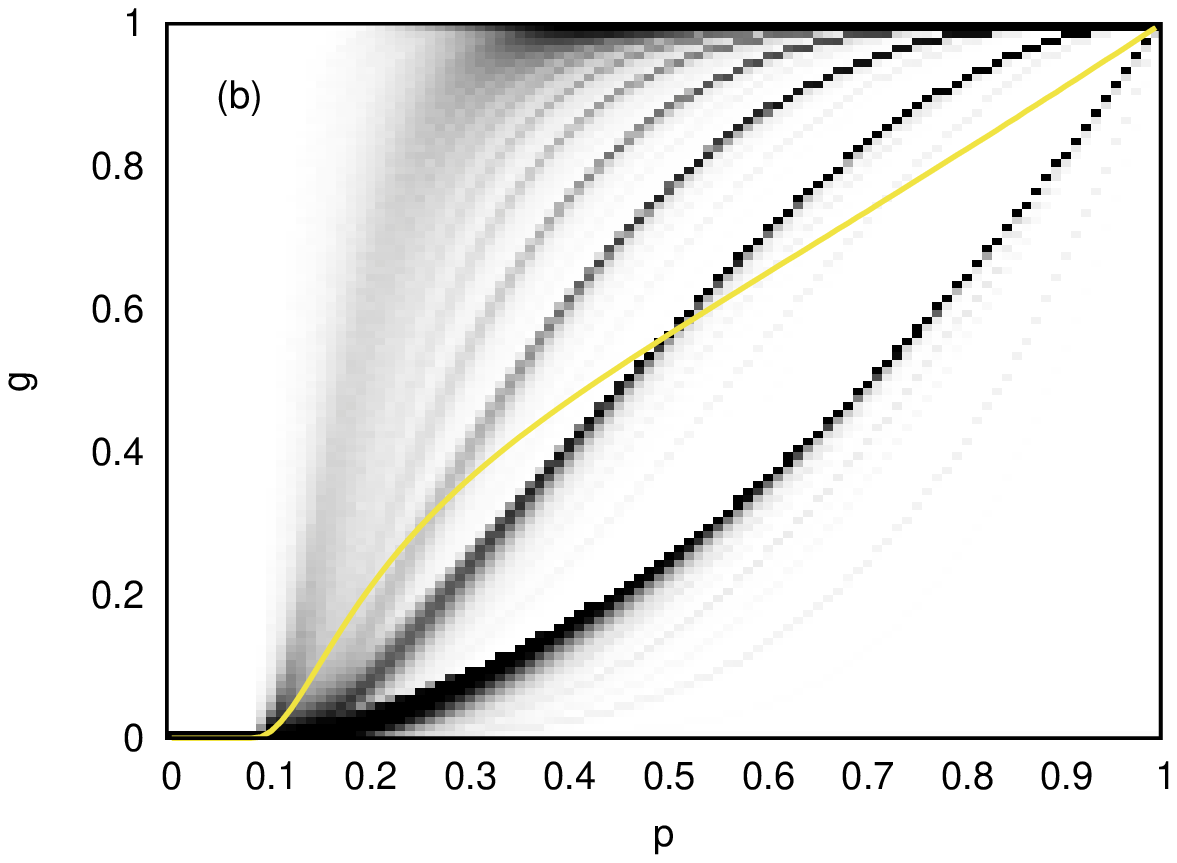}\\
\includegraphics[width=0.475\textwidth]{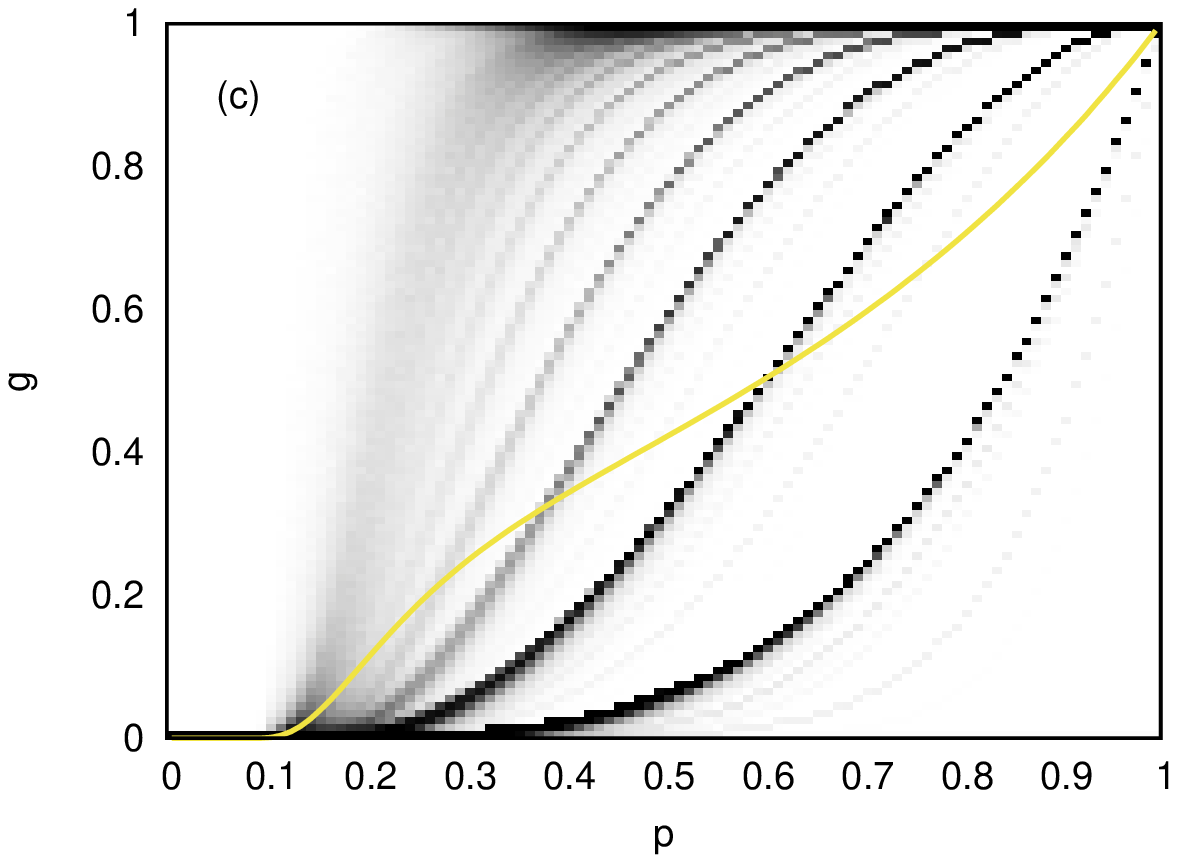}\hfil 
\includegraphics[width=0.475\textwidth]{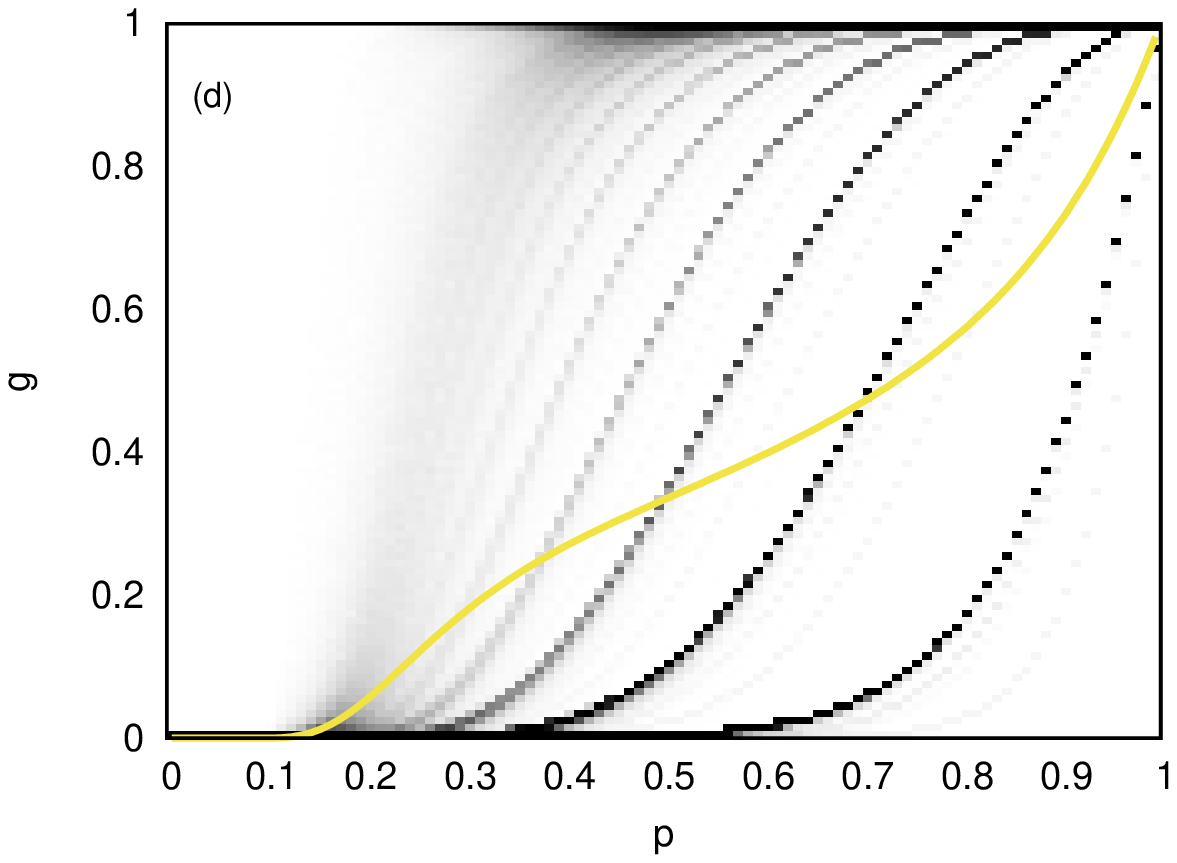}
\caption{
Distributions $\pi_\tau(g)$ of percolation probabilities for a sample network with $N= 62,586$ nodes, constructed as an undirected version of the Gnutella file sharing platform data. Grey scale coded probabilities are shown for all probabilities of retaining bonds with darker grey representing a higher probability. Also  shown are average percolation probabilities (full lines).  Panel (a) corresponds to 1 instance  of the percolation experiment, panel (b) to 2, panel (c) to 4  and panel (d) to 8 instances. Note that the upper left panel shows the data for a single percolation experiment also presented in Fig.~1 of \cite{KuRog17}.}
\label{GCStabGnutella}
\end{figure*}

In principle, distributions of any form of $n$-point correlations can be evaluated using the theory presented in Sects.\, III.A--D, for both large single instances of real world networks and synthetic network ensembles in the thermodynamic limit, provided they are in the configuration model class and are sparse with finite mean degree. However, it is clearly impossible to give a comprehensive overview of all results one {\em could\/} obtain, given the large number of parameters to play with such as degree distributions, bond retention probabilities, or the number of, and correlations between instances of the percolation process. Therefore, we illustrate some key aspects on a number of representative examples.

\begin{figure*}[t!]
\includegraphics[width=0.475\textwidth]{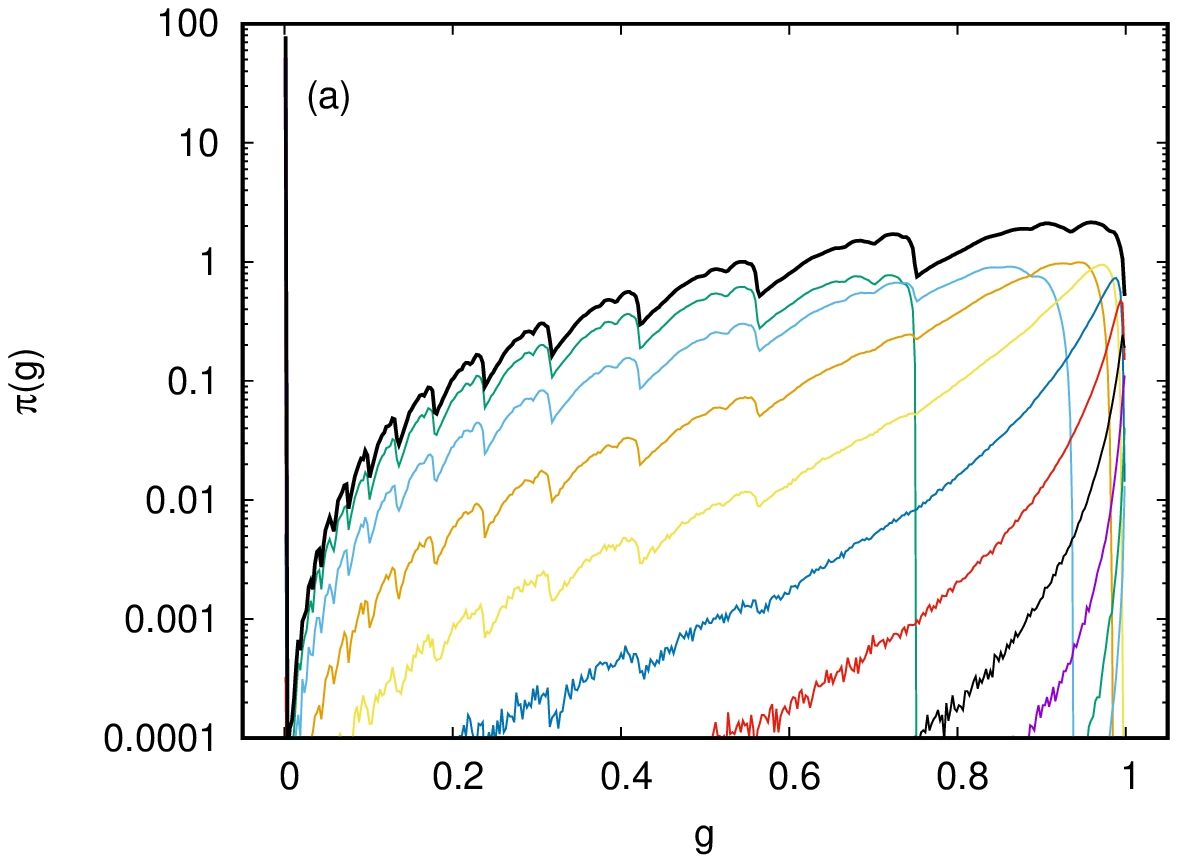}
\hfil 
\includegraphics[width=0.475\textwidth]{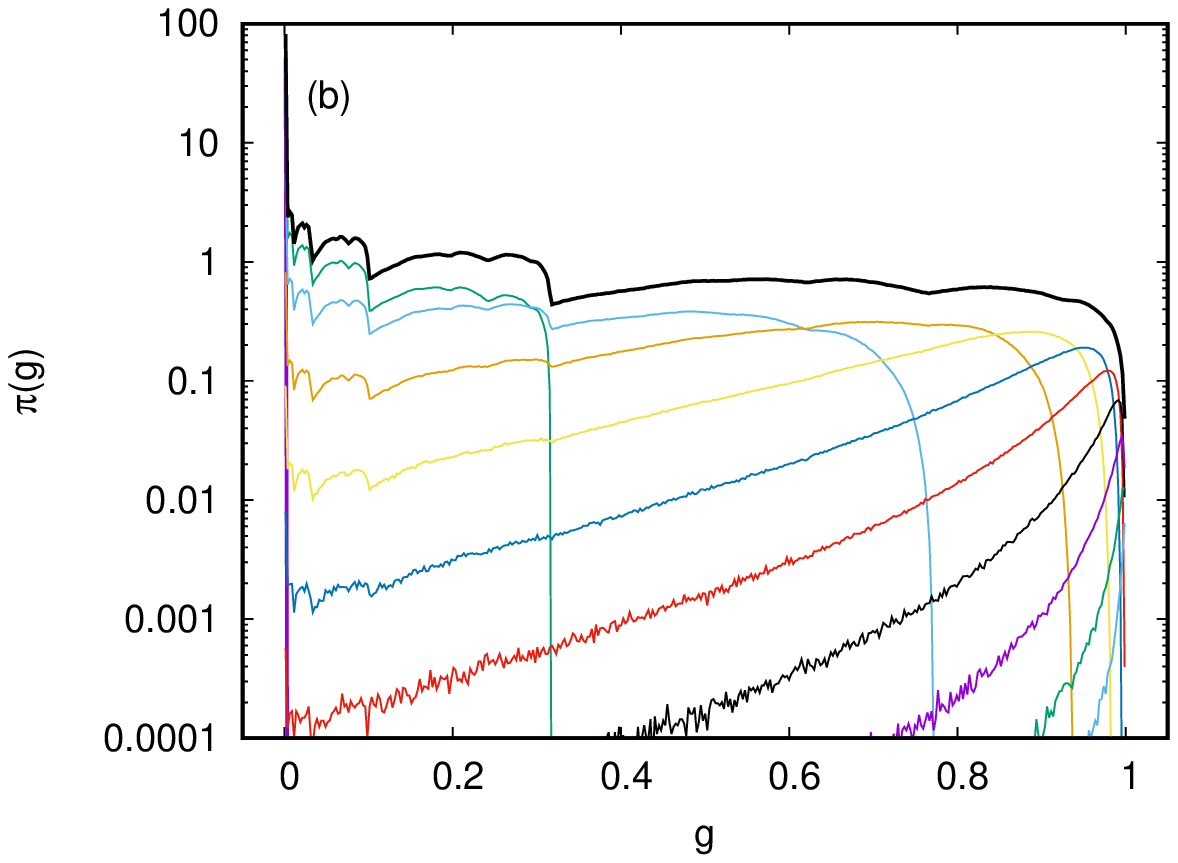}
\\ 
\includegraphics[width=0.475\textwidth]{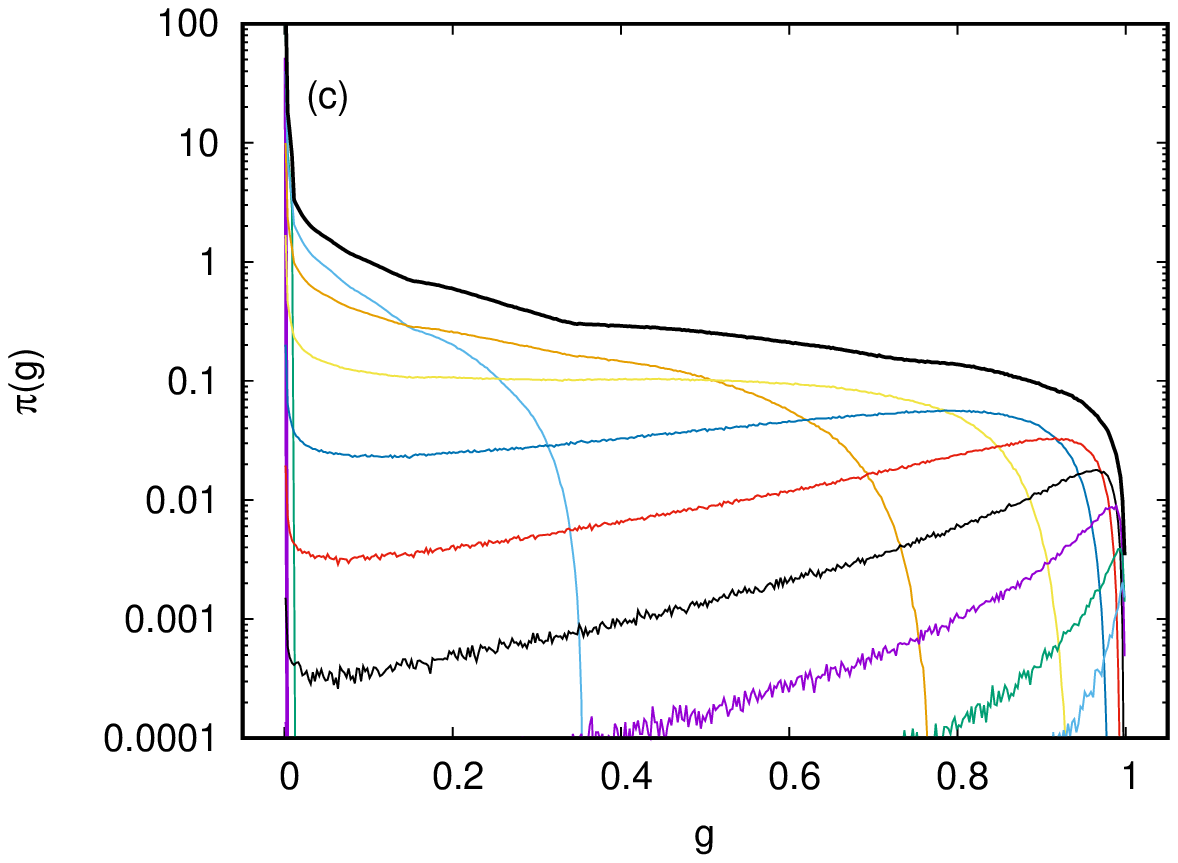}
\hfil 
\includegraphics[width=0.475\textwidth]{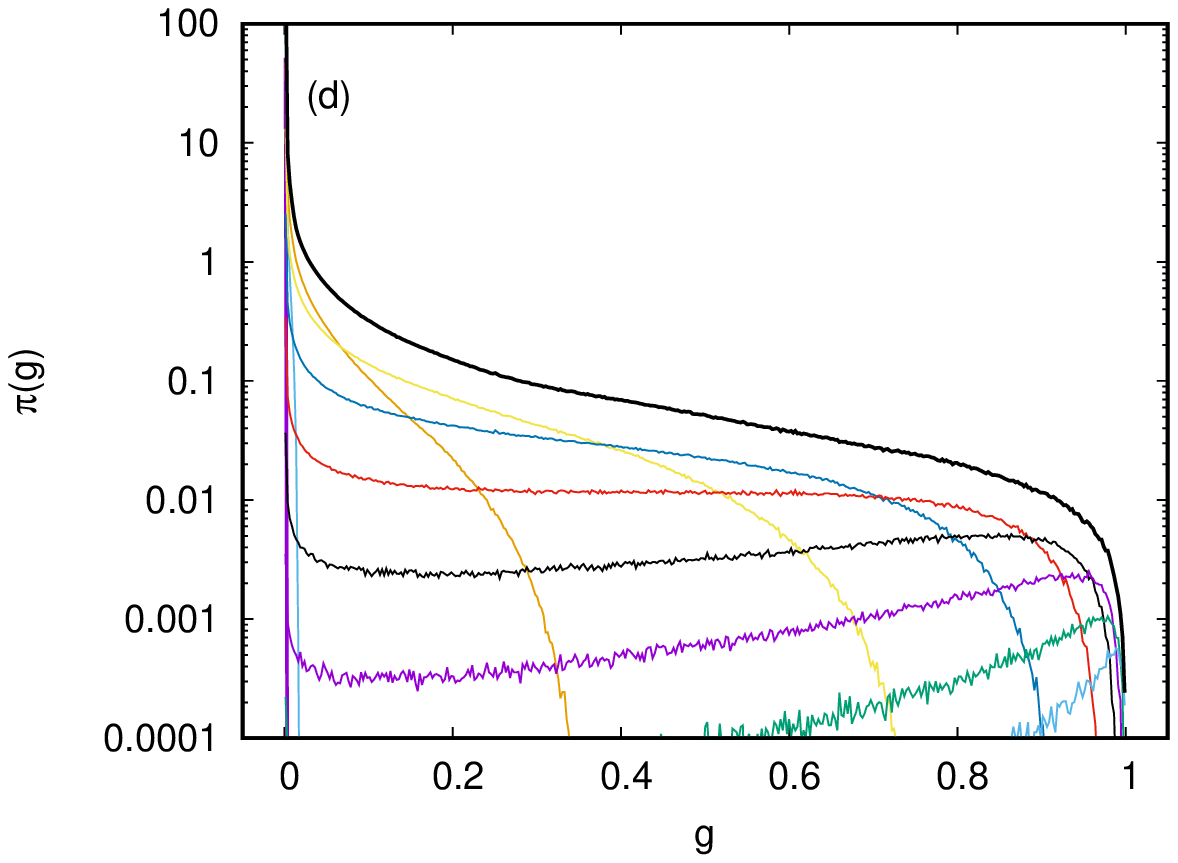}       
\caption{
Distributions $\pi_\tau(g)$ of percolation probabilities for an ER network of mean degree $c=2$, for repeated instances of percolation, with bond retention probability $p=0.75$ (top full lines), and their deconvolutions according to degree, i.e., the individual  contributions $p_k \pi_\tau(g|k)$ to the total, for $k=0, 1,\dots, 9$ and $k\ge 10$. The $\delta$-peaks at zero in each panel corresponds to $p_0 \pi_\tau(g|0)$. Distributions of percolation probabilities of nodes with higher degree are, not unexpectedly, supported at larger $g$. Panels (a)--(d) correspond to $\tau = 1, 4, 16$, and 64 instances of the percolation experiment, respectively.}
\label{GCStabER075}
\end{figure*}
\twocolumngrid

We first present the outcomes of multiple percolation instances at a local level, using an undirected version of the Gnutella file sharing platform data \cite{Lesk14} as example. In Fig.\, 1 we show grey-scale coded probability density functions (pdfs) of joint percolation probabilities for 1, 2, 4, and 8 instances of a percolation experiment as a function of the bond retention probability $p\in[0,1]$. Given the large range of values of the pdfs, they have been non-linearly transformed into a grey-scale mapping with darker tones corresponding to larger probabilities. In each panel we also show the {\em mean} probability for nodes to be in the GC for all instances of the percolation process. It is clear that the mean probability (the first moment of the pdf shown in grey-scale code), are only a very coarse reflection of the heterogeneity of outcomes in this problem. Notable in all panels are the distinct bands, with (in the case of the Gnutella data) fairly sharply defined upper edges, that become more and more blurred as we approach the critical percolation probability $p_c$ from above. We further note that the single instance result in the upper left panel replicates a result of \cite{KuRog17}. 

The main bands correspond to the contribution to $\pi_\tau(g)$ from nodes of different degrees $k$. The location of the sharp upper cut-offs of the main bands can be predicted from Eqs.\,\eqref{pitau-g}, \eqref{pitau-gk}, as $0\le\tilde g_\nu\le 1$. Upon insertion of the upper bound $\tilde g_\nu =1$, one obtains an upper bound for the support of the contribution of degree-$k$ nodes to $\pi_\tau(g)$, viz. $g\le \big(1-(1-p)^k\big)^{|\tau|}$. For $p$ sufficiently far above the percolation threshold, this agrees very well with the data in all panels. As expected, the locations of the bands move to lower values of $g$ with increasing number of instances.

Other less prominently visible bands can also be predicted by considering properties of the first coordination shell around a vertex. Formally this is done by replacing each of the $\tilde\pi(\tilde g_\nu)$ appearing on the r.h.s. of  Eq.\,\eqref{pitau-gk} by its expression in terms of the r.h.s. of the self-consistency equations Eq.\,\eqref{pit-gt}, \eqref{pit-gtk} and applying the same logic concerning the range of values for the $\tilde g_{\nu'}$ contributing to each $\tilde\pi(\tilde g_\nu)$. This process can be iterated to rationalize finer and finer details in the distributions. It is worth noting that the reasoning regarding the location of bands and their cutoffs is {\em independent\/} of the degree distribution. Thus bands with cutoffs will be visible in such representations for {\em  any} network. However, the band edges may be less pronounced since they will be clearly identifiable only if there is a sufficiently high density of cavity probabilities $\tilde g$ close to their upper cutoff. As a general rule, therefore, sharply defined bands and band edges will be observable only sufficiently far above the percolation threshold. Indeed, as the percolation threshold is approached typical values of the $\tilde g_\nu$ will be smaller than 1, entailing that the bands in Fig.\, 1  blur and start to overlap. To illustrate this more quantitatively, we show results for pdfs of joint percolation probabilities for two synthetic network ensembles at fixed bond retention probability $p$ in Figs.\, 2 and 3. Note that each figure  would correspond to a vertical cut at fixed $p$ in the representation chosen in Fig.\, 1.

Figure \ref{GCStabER075} shows the pdfs $\pi_\tau(g)$ for joint percolation probabilities in multiple instances of the percolation process, for an ER network of mean degree $c=2$ with a bond retention probability $p=0.75$. Results are obtained in the thermodynamic limit solving Eqs.\,\eqref{pit-gt}, \eqref{pit-gtk}, and using the solution to evaluate Eqs.\,\eqref{pitau-gk} and \eqref{pitau-g}. Panels correspond to results for different numbers of instances of the percolation process. Along with the full pdfs we also show their degree-based deconvolutions defined by Eqs.\,\eqref{pitau-g} and \eqref{pitau-gk}. Band edges appear far less sharp than for the Gnutella network, in part as contributions from different degrees strongly overlap. Nonetheless, in the upper left panel (single instance), the upper band edges from degree 1 nodes at $g=0.75$ and degree 2 nodes at $g=0.9375$ can be clearly discerned. Also prominent are satellite sub-band edges due to the degrees of nearest neighbours of degree 1 sites and to combinations of degrees of the neighbours of degree 2 sites. Again the dominant weight of joint percolation probabilities and the various band edges move to lower values of $g$ with increasing number of instances. For example (upper right panel in Fig.\,\ref{GCStabER075}.), with four uncorrelated instances the upper band edge for degree 1 nodes is at $g=p^4\simeq 0.3164$, and at $g=(1-(1-p)^2)^4 \simeq 0.7725$ for degree 2 nodes.

\begin{figure*}[t!]
\includegraphics[width=0.475\textwidth]{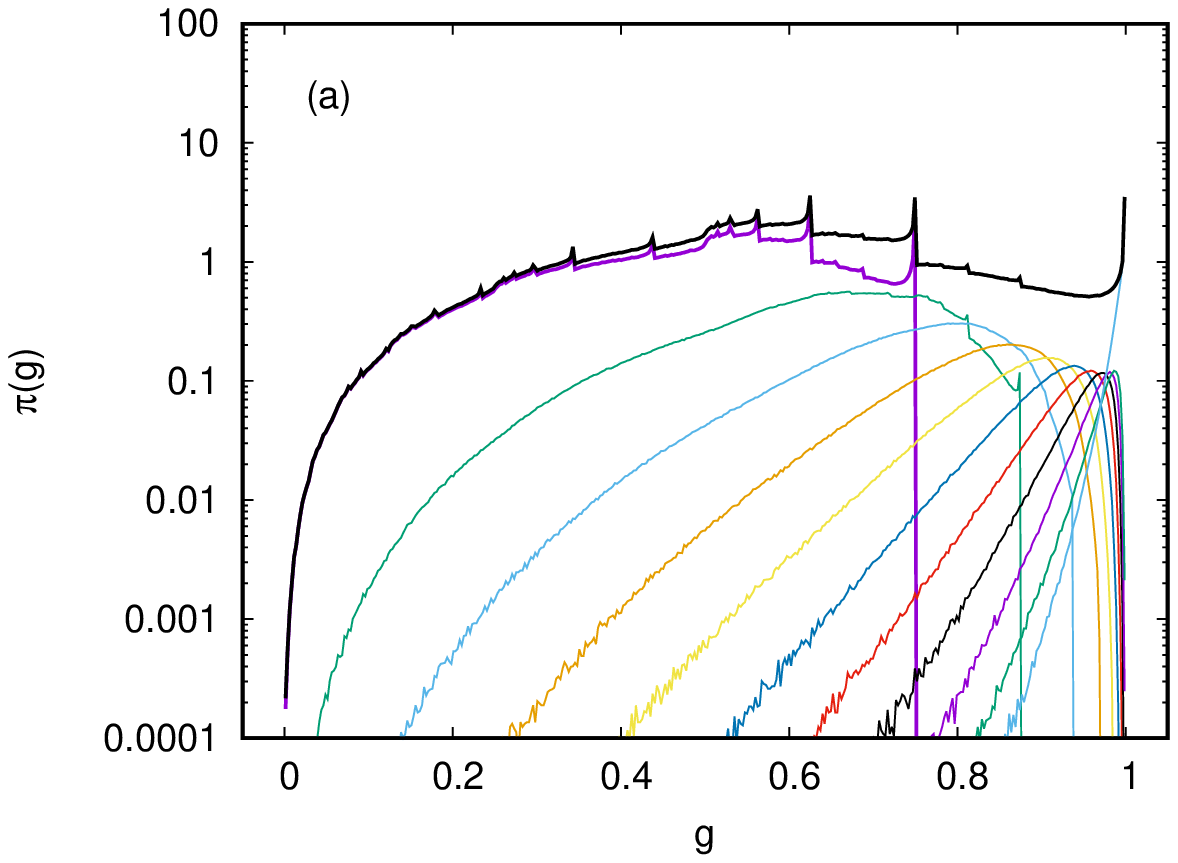}
\hfil 
\includegraphics[width=0.475\textwidth]{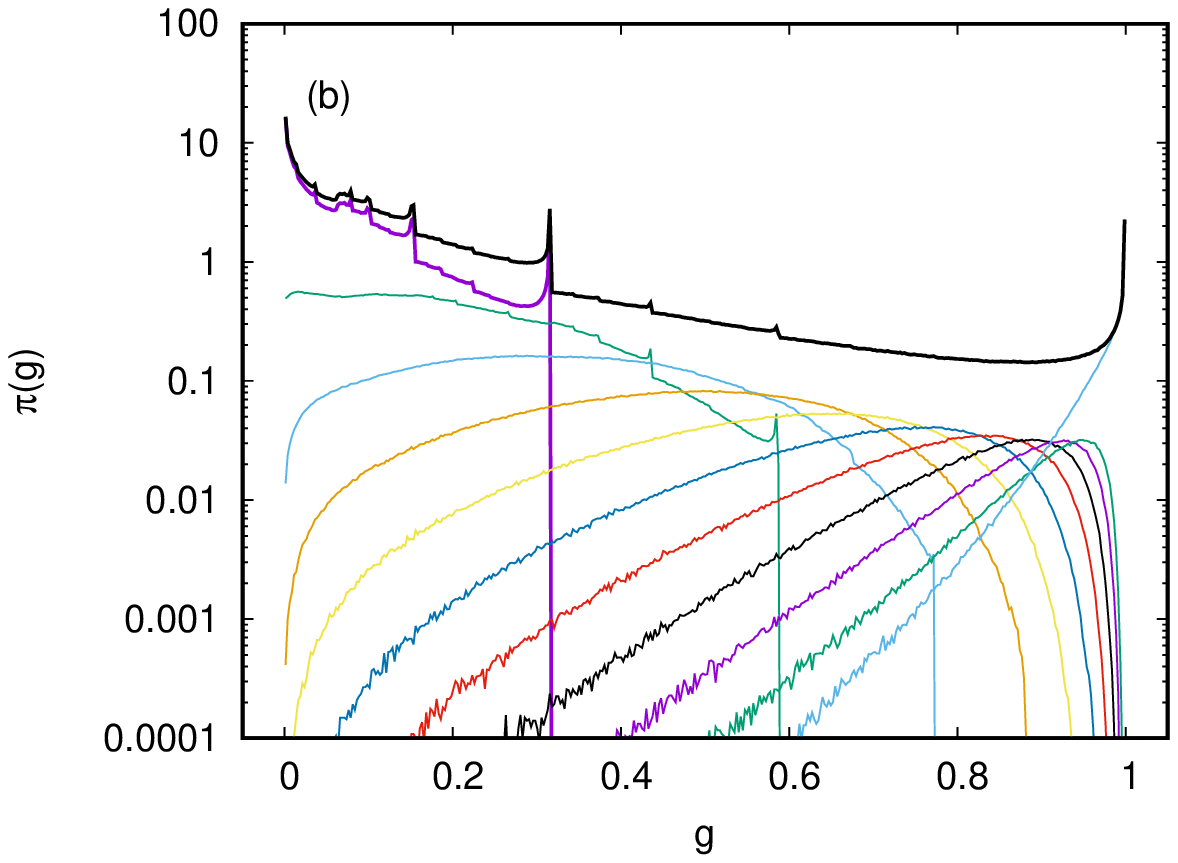}
\\
\includegraphics[width=0.475\textwidth]{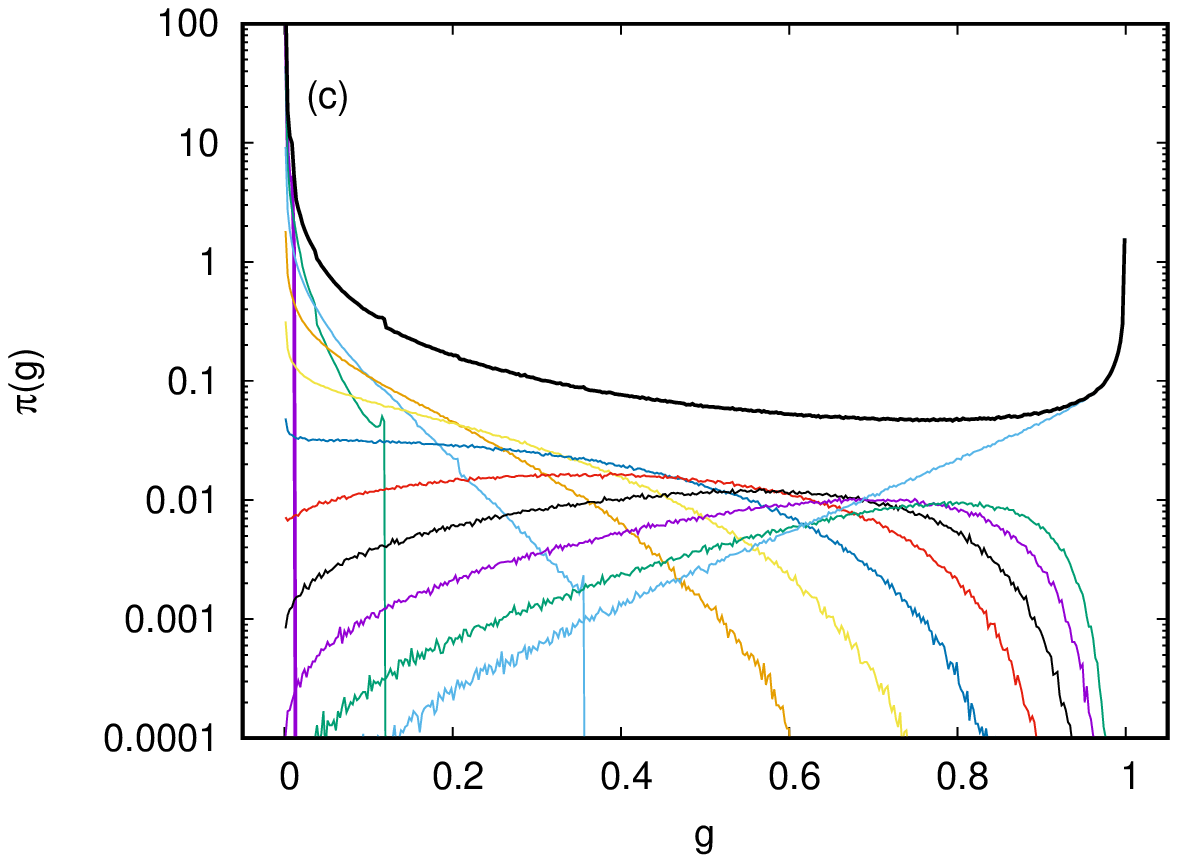}
\hfil 
\includegraphics[width=0.475\textwidth]{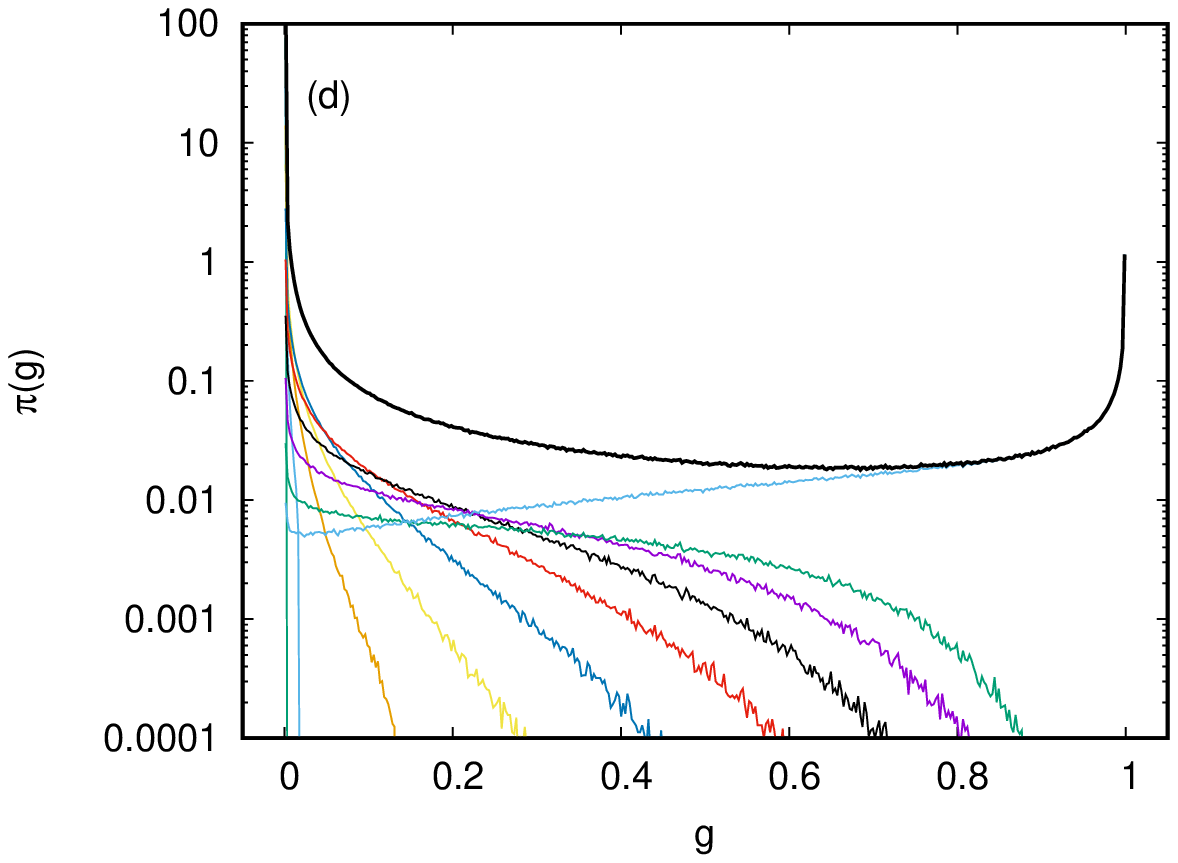}	       
\caption{
 Distributions $\pi_\tau(g)$ of percolation probabilities for a scale free network with degree distribution $p_k \propto k^{-3}$, $k\ge 2$, for repeated instances of percolation,  with bond retention probability $p=0.5$ (top full lines), and their deconvolution according to degree for $k=2, 3,\dots, 11$ and $k\ge 12$. As in the ER case, distributions of percolation  probabilities of nodes with higher degree are supported at larger $g$. Panels (a)--(d) correspond to $\tau = 1, 4, 16$, and 64 instances of the percolation experiment, respectively.}
\label{GCStabpow3m205}
\end{figure*}

Figure \ref{GCStabpow3m205} shows analogous results for a scale-free network with degree distribution $p_k \propto k^{-3}$, for $k\ge 2$, but now with $p=0.5$ as the bond retention probability, again evaluated in the thermodynamic limit using Eqs.\,\eqref{pit-gt}, \eqref{pit-gtk} and \eqref{pitau-g}, \eqref{pitau-gk}. Band edges are more sharply defined in this case, mainly because the large weight of events with $\tilde g \to 1$. Their location, however, is the same as in the Gnutella example and as it would be in an ER network at the same value of $p$. A main noticeable difference compared to the ER network is the survival of the peak of the pdf of joint percolation probabilities at $g \lesssim 1$ even for large numbers of repetitions of the percolation process. As already noted in \cite{Kitsak+18}, this is due to the presence of hubs with high degrees in systems with broad degree distributions. Here we can quantitatively confirm this at a local level: as shown in the lower right panel of Fig.\,\ref{GCStabpow3m205}, for 64 instances virtually all contribution to the joint pdf of percolation probabilities at $g\ge 0.5$ comes from nodes with degrees $k\ge 12$.

\begin{figure*}[t!]
\includegraphics[width=0.475\textwidth]{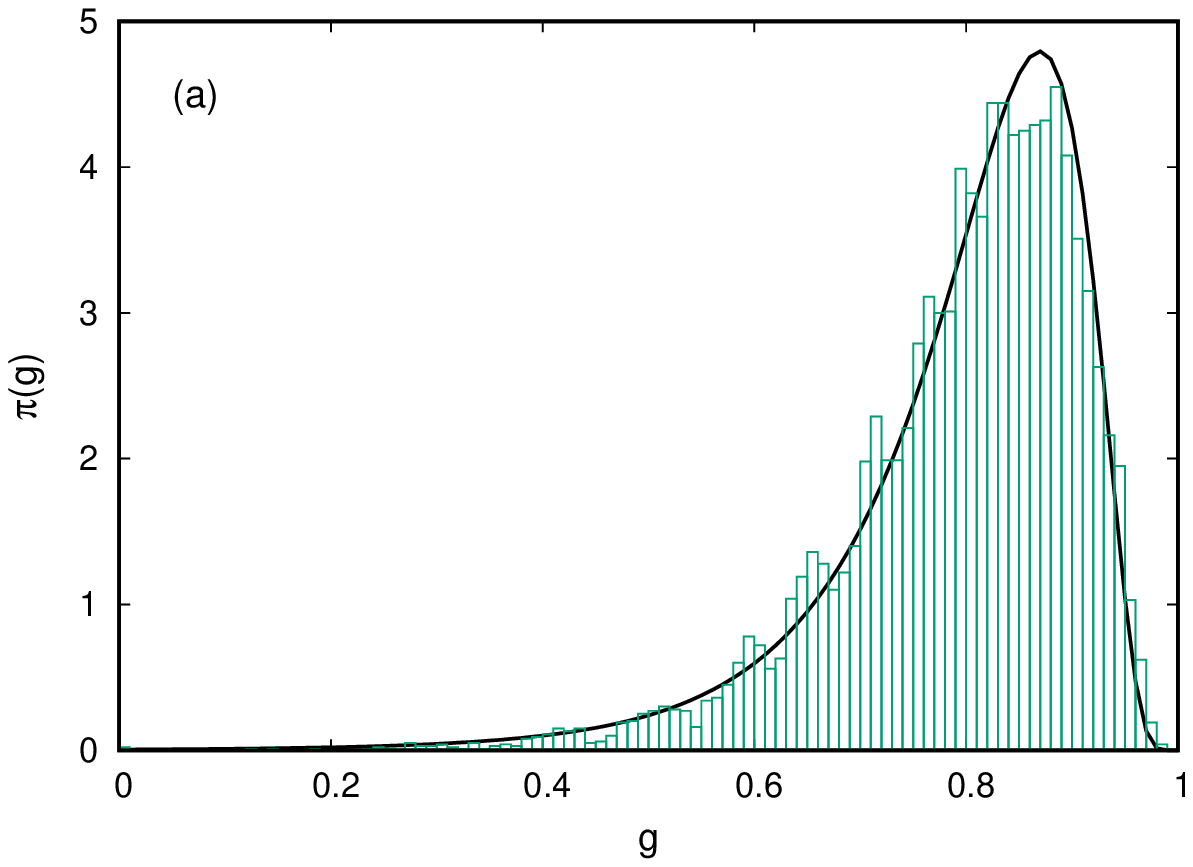}\hfil 
\includegraphics[width=0.475\textwidth]{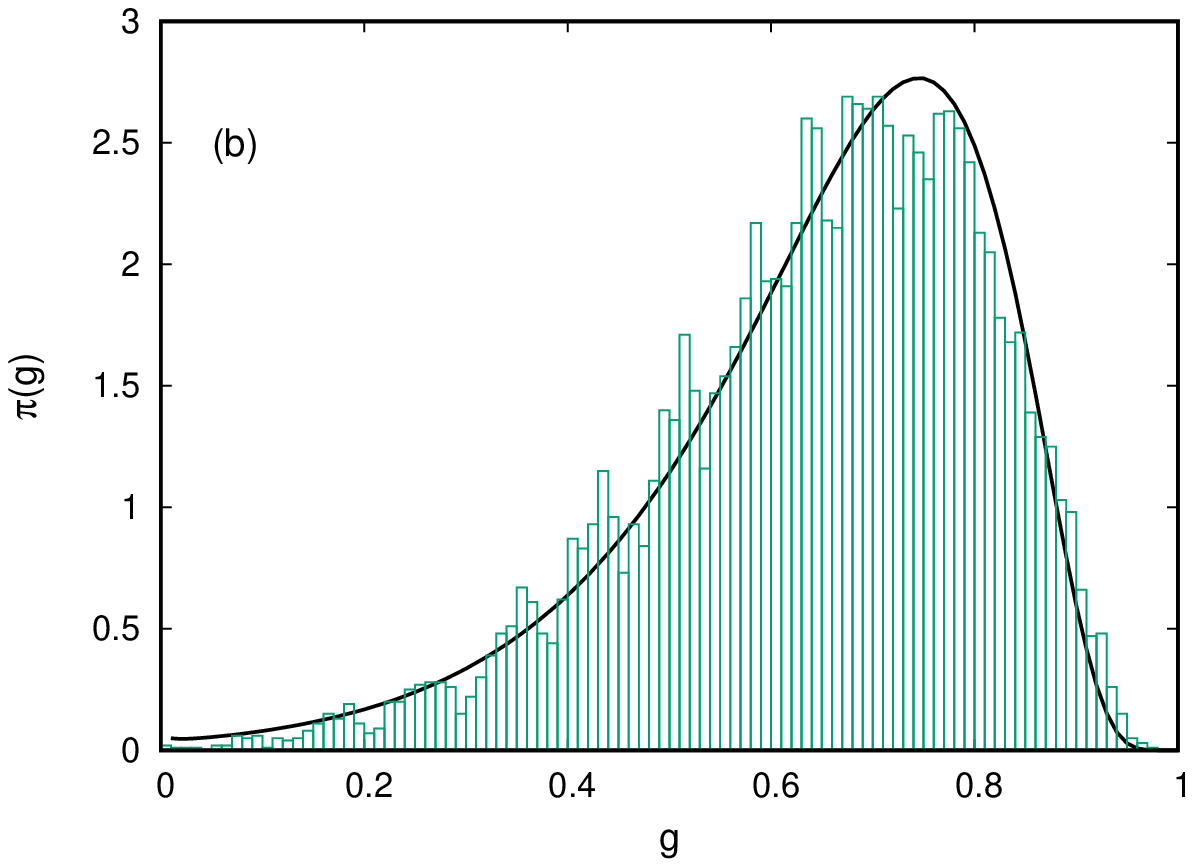}\\
\includegraphics[width=0.475\textwidth]{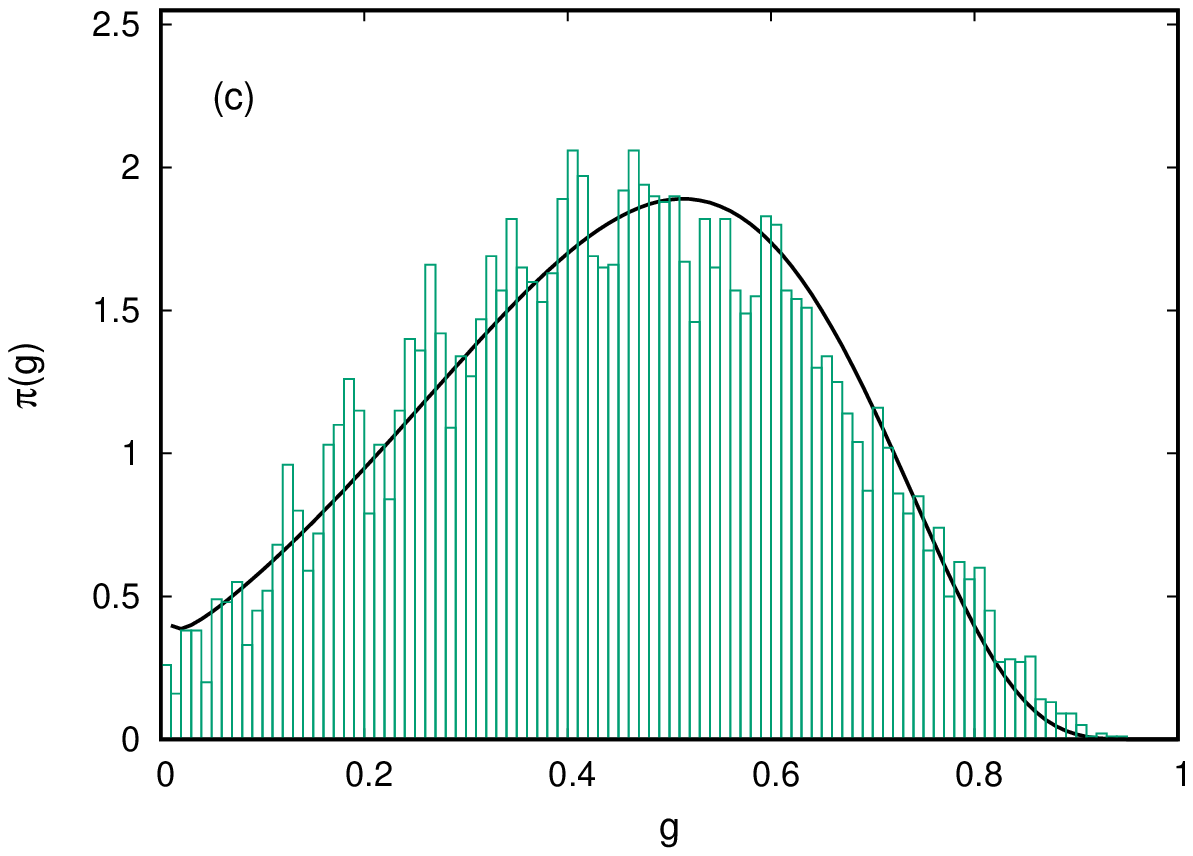}\hfil 
\includegraphics[width=0.475\textwidth]{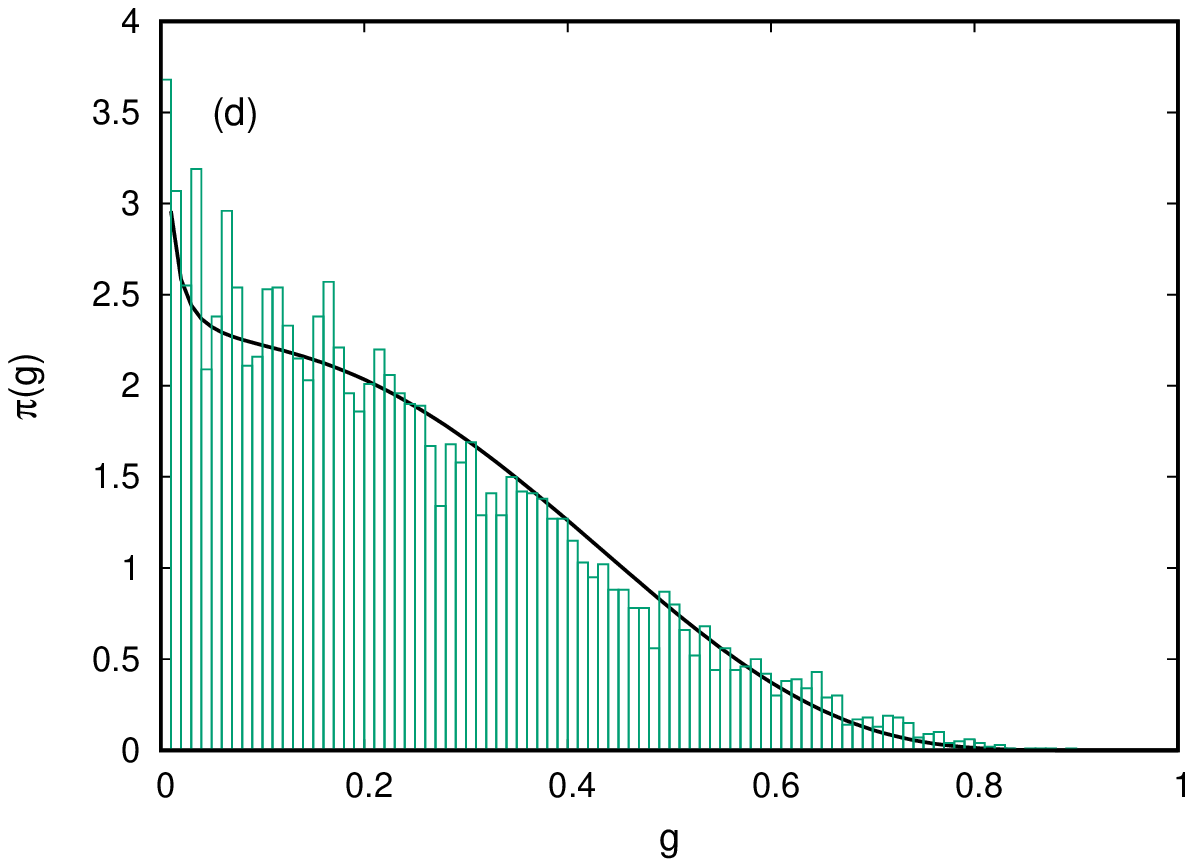}\\
\caption{
Distributions $\pi_\tau(g)$ of joint percolation probabilities for an ER network of mean degree $c=10$, comparing results of the large mean degree approximation Eq.~\eqref{piofg-largec}  (black full line) with results obtained for a large single realization of a network with $N=10,000$ nodes (histograms) with bond retention probability $p=0.2$. Panels (a)--(d) correspond to 1, 2, 4  and 8 instances, respectively.}
\label{GCStabER10}
\end{figure*}

Fig.\,\ref{GCStabER10} compares results of the large mean degree approximation Eq.~\eqref{piofg-largec} for an ER network with results obtained for a single such network of $N=10,000$ nodes, with mean degree $c=10$, for $|\tau| =1, 2, 4$, and $8$ uncorrelated instances with bond retention probability $p=0.2$. Despite the moderate value of the mean degree, the main features of these distributions are captured fairly well by the large mean degree approximation. Nevertheless, it misses some of the fine structure of well identifiable peaks which are due to  different types of local environment of nodes.

\begin{figure*}[t!]
\includegraphics[width=0.475\textwidth]{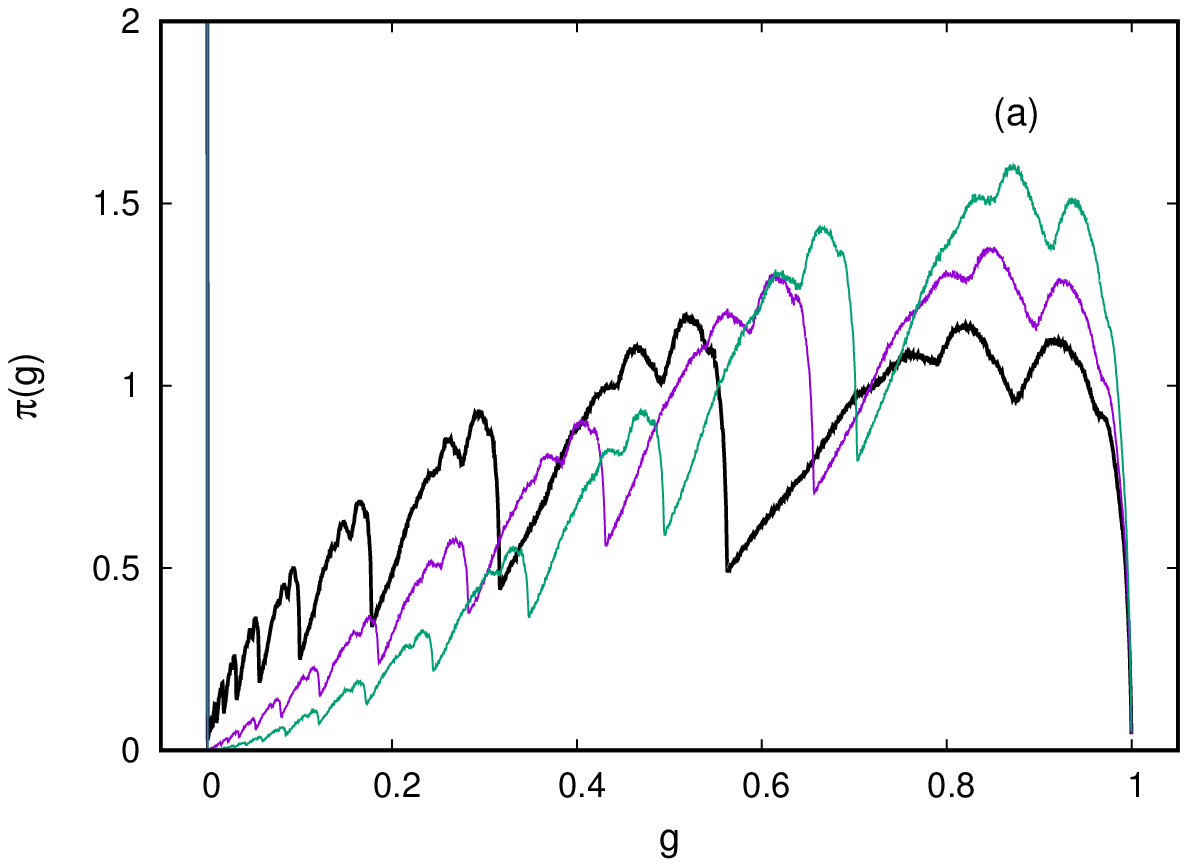}
\hfil 
\includegraphics[width=0.475\textwidth]{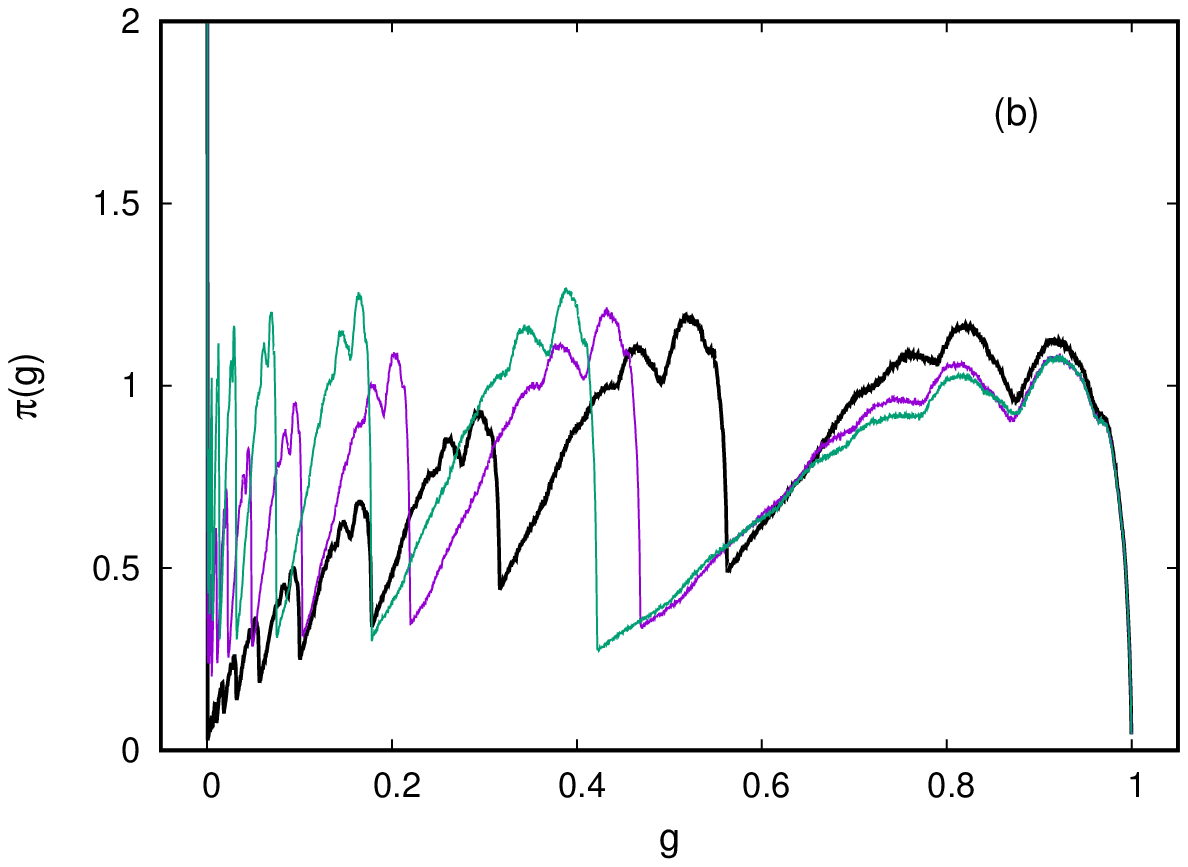}
\\
\includegraphics[width=0.475\textwidth]{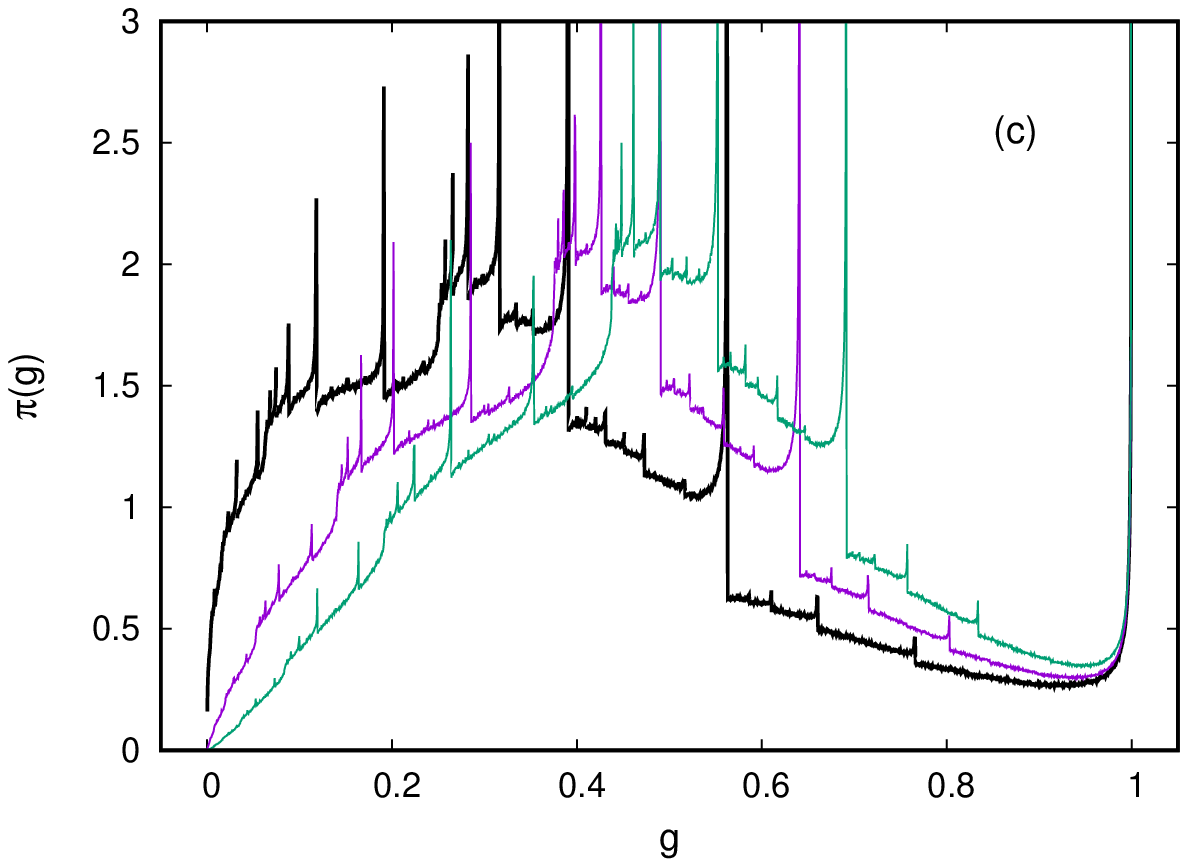}
\hfil 
\includegraphics[width=0.475\textwidth]{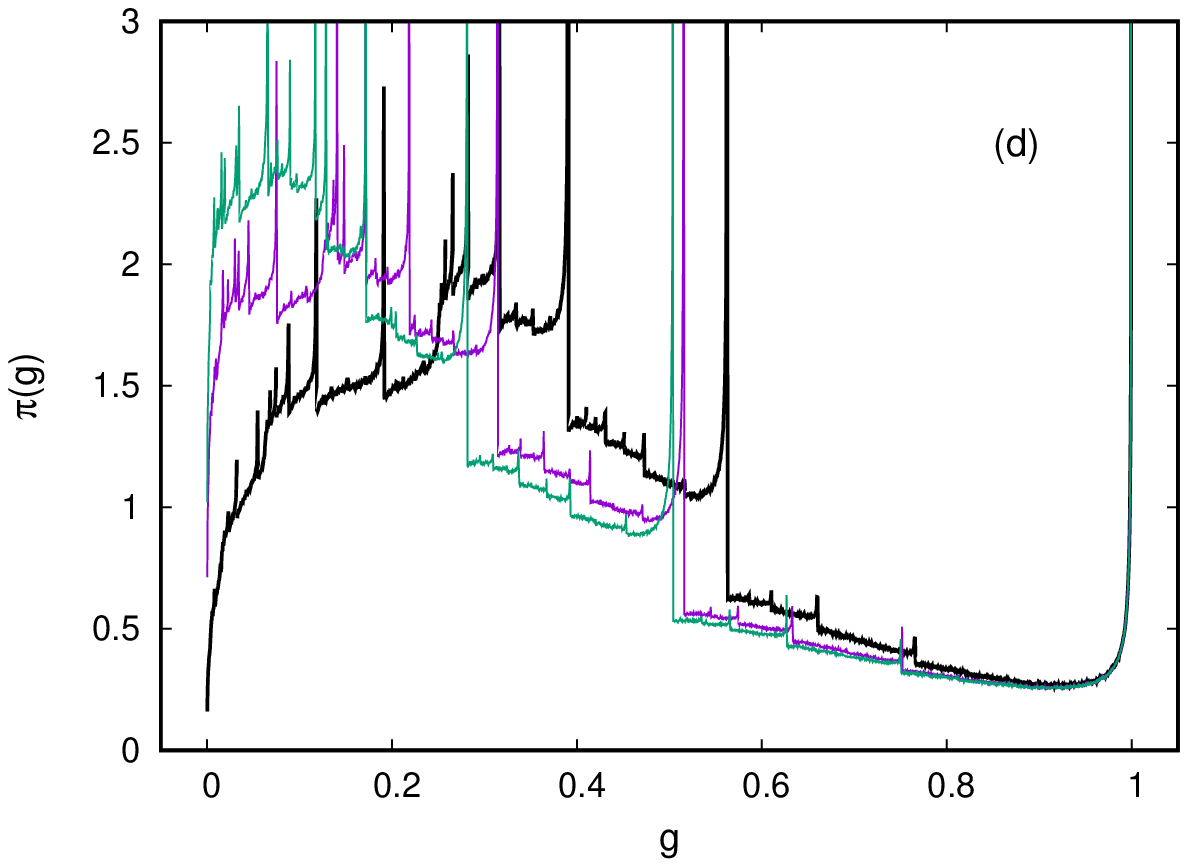}
\caption{
  Distributions $\pi_2(g)$ of joint percolation probabilities for the ER network at $p=0.75$ considered in Fig.\,\ref{GCStabER075} (panels (a) and (b)) and for the scale free network at $p=0.5$ considered in Fig. \ref{GCStabpow3m205} (panels (c) and (d)) for two instances of the percolation process. Panels (a) and (c) compare results for uncorrelated instances $r=0$ (black full line) with positively correlated instances with coefficients $r=0.5$ (purple line) and $r=0.75$ (green line). Panels (b) and (d) compare results for uncorrelated instances $r=0$ (black full line) with  negatively correlated instances with coefficients $r=-0.5$ (purple line) and $r=-0.75$ (green line). }
\label{GCStab2Corr}
\end{figure*}

In Fig.\,\ref{GCStab2Corr} we plot the pdfs of joint percolation probabilities to illustrate the effect of correlation between two instances. Panels (a) and (b) row show results for the ER network considered in Fig.\,\ref{GCStabER075} (mean degree $c=2$, bond retention probability $p=0.75$), while  panels (c) and (d) show results for the scale free network considered in Fig.\,\ref{GCStabpow3m205} (degree distribution $p_k \propto k^{-3}$, for $k\ge 2$, bond retention probability $p=0.5$).  Left and right panels compare uncorrelated instances with positively and negatively correlated ones respectively at different values of the (anti-)correlation coefficient. 

We note that negative correlation increases the probability of having low values of the joint percolation probability and suppresses the probability of having large values of the joint percolation probability.  The opposite trend is observed for positive correlation. These results are plausible as negative correlation in the percolation process enhances the probability to remove {\em different\/} edges in the two instances, thus decreasing the likelihood that nodes remain on the GC in {\em both} instances. Conversely, positive correlation increases the probability to remove the {\em same\/} edges in the two instances, thus increasing the likelihood that (the same) nodes remain on the GC in {\em both} instances. In the extreme case where the correlation approaches 1, the joint effect of two highly correlated instances becomes indistinguishable from that of a single instance.

\begin{figure*}[t!]
\includegraphics[width=0.475\textwidth]{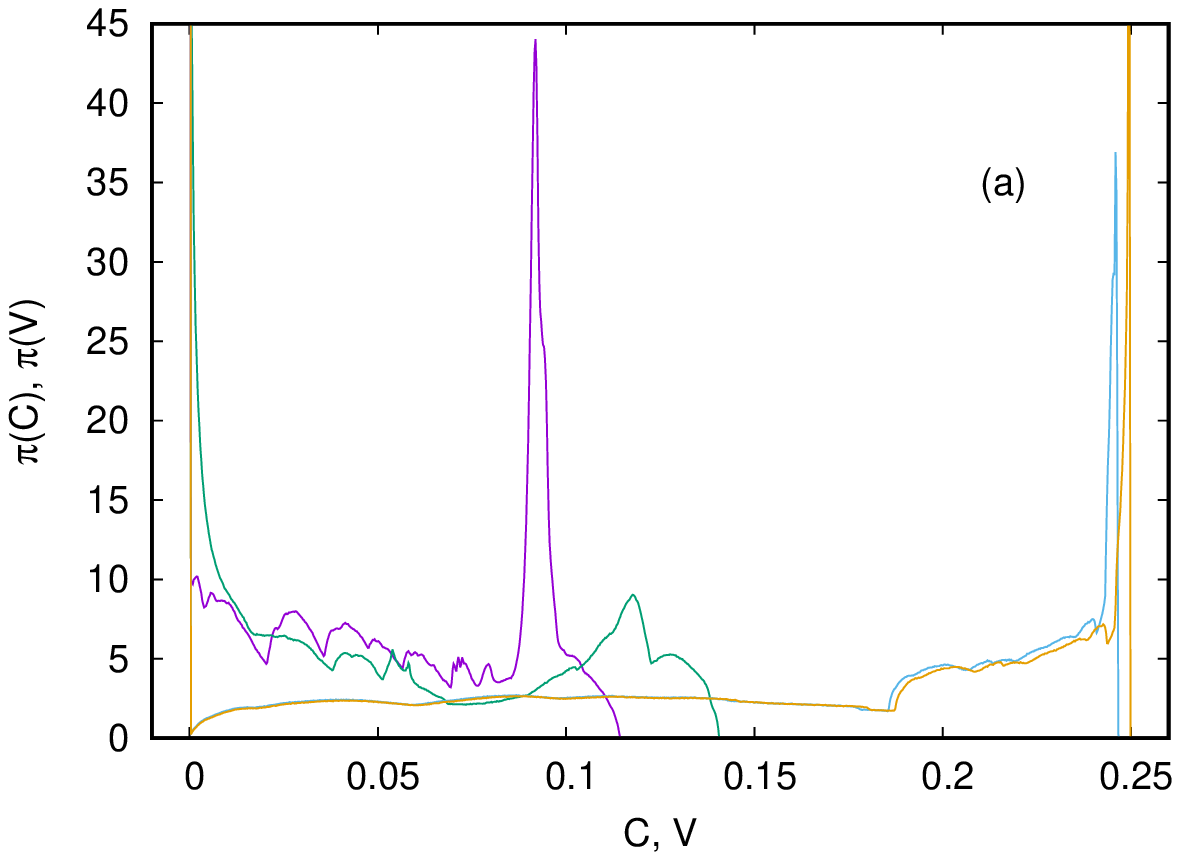}\hfil
\includegraphics[width=0.475\textwidth]{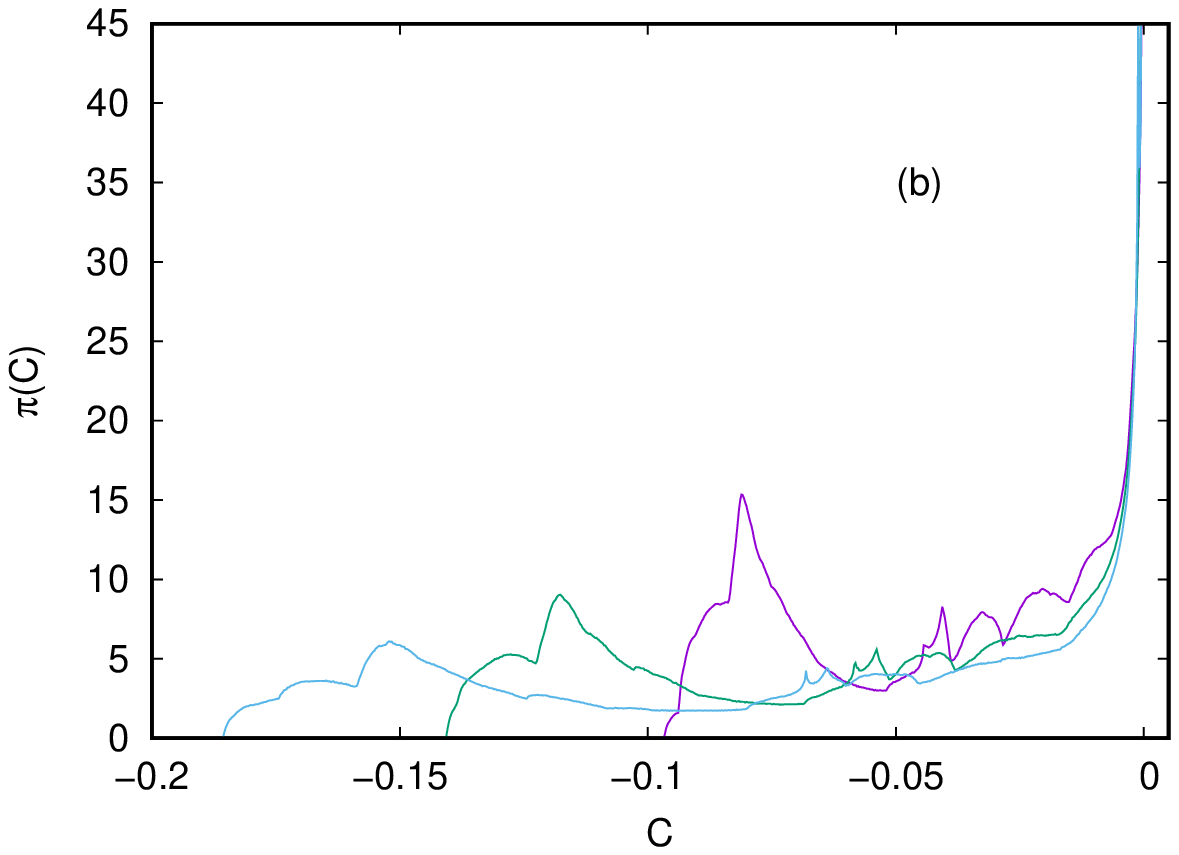}
\caption{Panel (a):   Distribution $\pi(C)$ of the co-variances between two instances of a percolation process on an ER network of mean degree $c=2$,  with bond retention probability $p=0.75$ with correlation coefficients $r=0.5, 0.75$, and $0.99$ (narrow to broad curves), respectively. Also shown is the single instance distribution $\pi(V)$  of variances of percolation probabilities. The distribution of co-variances for the  highly correlated case is very close to the distribution of variances, which exhibits a divergence at $V=0.25$. Panel (b): Distribution $\pi(C)$ of the co-variancec between two instances of a percolation process for the same system, but for {\em anti-correlated\/} instances, with correlation coefficients $r=-0.5, -0.75$, and $-0.99$ (narrow to broad curves), respectively.
}
\label{GCStabERs2Cov}
\end{figure*}

Finally, in Fig.\,\ref{GCStabERs2Cov} we plot the distribution of the co-variances  $C_i=\langle n_i(t)n_i(t')\rangle-\langle n_i(t)\rangle\langle n_i(t')\rangle$ for pairs of correlated instances, taking the ER network (mean degree $c=2$,  bond retention probability $p=0.75$) as an example. The results demonstrate once more that there is rich structure in the distribution of the co-variance, which the average as evaluated in \cite{Bianconi17} cannot reveal. The left panel of Fig.\,\ref{GCStabERs2Cov} shows distributions of the co-variances for several positive values of correlation between the two instances. As $r\to1$, we expect the distribution of the co-variances between two instances to approach that of the variance of a single instance. Comparison of the distribution of the co-variances at $r=0.99$ with the distribution of the variances in Fig.\,\ref{GCStabERs2Cov} shows this indeed to be the case. The right panel of Fig.\,\ref{GCStabERs2Cov} shows the distributions for several negative values of correlation between the two instances. Clearly the distribution of the co-variances is now supported at negative values, but the distributions of the co-variances is {\em not\/} just the mirror image of the corresponding distribution at positive $r$. This will only be the case for a value $p=0.5$ of the bond retention probability. For the covariances, a deconvolution by degree reveals that, e.g. in the $r=0.5$-case, the dominant contribution to the peak at $C\simeq 0.092$ comes from the $k=1$ sites, while most of the structure at smaller $C$ is due to $k=2$ sites. Analogous features are observed in the anti-correlated case (e.g. at $r=-0.5$), where the the dominant peak at $C\simeq -0.081$ is mainly due to $k=1$-sites, and further structures at smaller $|C|$ are mainly due to $k=2$-sites. High degree sites will in general only show \textit{small} values of $C$ (or $|C|$ in the anti-correlated case).

\section{Summary and Discussion} \label{sec:SumDisc}
In the present paper, we have analyzed the heterogeneity of outcomes of percolation in complex networks in a set of several possibly correlated instances of the percolation process, thereby expanding the analysis of \cite{KuRog17} to this more complex problem. At a global level these problems were recently considered in \cite{Bianconi17} with emphasis on average variance and average co-variances in pairs of instances, and in \cite{Kitsak+18} with emphasis on average joint percolation probabilities for multiple instances of a percolation process. The latter were advocated as a measure of the stability of the giant (or percolating) component in a given network. The problem is clearly relevant when assessing the robustness of the functionality of supply or communication infrastructures against repeated failures of components.

With that context in mind it becomes clear, however, that average joint percolation probabilities are not necessarily the most appropriate measure, and that full distributions of joint probabilities contain far more information regarding the exposure of key components of a net against repeated failures of components elsewhere in a net. This was the main reason to embark on the present project. One would in particular want to ensure for critical components in a network to be connected in such a way that the probability for them to remain part of the GC even in many instances of a percolation process remains close to 1.

Although in the present paper we have only considered the case of bond percolation, it would be straightforward to extend our analysis to node percolation or to percolation in directed networks.

Specifically, we have demonstrated that there is a considerable heterogeneity of the probabilities of individual nodes to remain part of the GC across instances of a percolation experiment, both for the Gnutella file-sharing network as an example of a real world network, and for synthetic networks in the configuration model class. While the degree of a node is an important feature influencing its joint percolation probability, it does not determine it entirely, as shown by the the fact that  the degree-dependent distributions of joint percolation probabilities are \textit{themselves} broad. The shape of these degree dependent distributions changes markedly with the number of percolation instances, and with the correlation between the instances. While positive correlation enhances large joint percolation probabilities and suppresses small joint percolation probabilities in comparison to uncorrelated instances, the opposite trend is observed for negative correlations.
We reiterate that the heterogeneity desribed in the present paper is different from that observed in explosive percolation, and that it is a {\em typical phenomenon\/} that is not caused by rare configurations of removed bonds.

The link between sizes of epidemics in a SIR models of infectious diseases and bond percolation allows us to expose the heterogeneity of the risk of being affected by a series of epidemics. Realistically, one should use different $p_t$ in such studies, and while this is covered by our general theory, we have produced results only for $p_t\equiv p$ for the sake of simplicity. The heterogeneity of risk-profiles exposed by our results might well be used for the design of vaccination strategies which would keep probabilities of infection for key personnel low across several epidemics.

In the present paper, we have not investigated any dynamic features of epidemic spreading, though these can be incorporated into a message passing approach as demonstrated in \cite{KarrerNewm10} for SIR models, and used to evaluate time-dependent average infection probabilities. A very interesting recent study \cite{MooreRog20} that  built on the results of \cite{KarrerNewm10} also reveals local dynamic features such as times to infection after outbreak. Studies of this type may be very useful to explore the efficiency of different social distancing strategies that might be contemplated in cases of a highly infectious epidemic, given that societies after implementation of distancing measures might be better described in terms of contact network structures than in terms of well-mixed populations (as assumed in classical mean-field theories). To be applicable to typical respiratory diseases, one would have to extend the message passing approach to capture the dynamics of a susceptible-exposed-infected-recovered (SEIR) model which is thought to better capture dynamic infection histories than the SIR model class, although SIR and SEIR models do exhibit the same epidemic threshold and the same asymptotic size as the SIR case. Also, in severe cases, where social distancing measures are indeed contemplated, transmission probabilities would be time-dependent and the effect of non-negligible fractions of recovered individuals at the time of an introduction of such measures (or, for that matter, their easing) would have to be taken into account in order to assess their effects. We believe that much of this is within reach of current techniques.
\bibliography{gc-stab.bbl}
\end{document}